\documentclass[11pt,a4paper]{article}
\usepackage{jheppub}
\usepackage{graphicx}
\usepackage{amsmath}
\usepackage{mathtools}
\input epsf
\usepackage{epsfig}
\usepackage{tikz}
\usepackage{subfigure,tikz}
\usetikzlibrary{backgrounds,automata}
\usepackage{amssymb}
\usepackage{graphics}
\usepackage{subfigure}
\usepackage[active]{srcltx}
\usepackage{amsthm}
\usepackage[T1]{fontenc} 
\usepackage{graphicx}

\newcommand{\bea}{\begin{eqnarray}}
\newcommand{\eea}{\end{eqnarray}}
\newcommand{\bean}{\begin{eqnarray*}}
\newcommand{\eean}{\end{eqnarray*}}
\newcommand{\nn}{\nonumber \\}

\def\abs#1{\left| #1\right|}

\def\eref#1{(\ref{#1})}

\def\d{{\rm d}}

\def\a{{\alpha}}

\def\b{{\beta}}

\def\d{\partial}

\def\eps{\epsilon}

\def\Label#1{\label{#1}%
  \smash{\hbox to0pt{\raise1ex\hbox{\tiny[#1]}\hss}}}

\newcommand{\parall}[2]{{#1}\ /\kern -0.8em / \  {#2}}

\preprint{USTC-ICTS/PCFT-21-30 }

\title{\boldmath One-loop Feynman Integral Reduction by Differential Operators}

\author[a,b]{Chang Hu,}
\author[a,1]{Tingfei Li,\note{Corresponding author.}}
\author[c,d]{Xiaodi Li}

\affiliation[a]{Zhejiang Institute of Modern Physics, Zhejiang University, Hangzhou, 310027, P. R. China }
\affiliation[b]{Hangzhou Institute of Advanced Study, UCAS, Hangzhou, 310027, P.R. China}
\affiliation[c]{Interdisciplinary Center for Theoretical Study, University of Science and Technology of China, Hefei, Anhui 230026, China}
\affiliation[d]{Peng Huanwu Center for Fundamental Theory, Hefei, Anhui 230026, China}

\emailAdd{isiahalbert@126.com}
\emailAdd{tfli@zju.edu.cn}
\emailAdd{xiaodili@ustc.edu.cn}

\abstract{For loop integrals, the standard method is reduction. A well-known reduction method for one-loop integrals is the Passarino-Veltman reduction. Inspired by the recent paper~\cite{Feng:2021enk} where the tadpole reduction coefficients have been solved, in this paper we show the same technique can be used to give a complete integral reduction for any one-loop integrals. The differential operator method is an alternative version of the PV-reduction method. Using this method, analytic expressions of all reduction coefficients of the  master integrals can be given by algebraic recurrence relation easily. We demonstrate our method explicitly with several examples.

}

\keywords{One-loop Feynman integral, Integral Reduction, Differential Operator, Recursion Relation}

\begin{document}

\maketitle

\flushbottom
\section{Introduction}

The calculation of scattering amplitude at higher loop level is  always like a chronic disease to block the evolution of High Energy Physics. Theoretical physicists have made many prescriptions to cure this problem started in the 1970s. The most significant receipt is to reduce a loop amplitude into a linear combination of some scalar master integrals (MIs) under dimensional regularization~\cite{ Brown:1952eu,Melrose:1965kb,Passarino:1978jh,tHooft:1978jhc,vanNeerven:1983vr, Stuart:1987tt,vanOldenborgh:1989wn,Bern:1992em,Bern:1993kr,Fleischer:1999hq,Binoth:1999sp,Denner:2002ii,Duplancic:2003tv,Denner:2005nn,Ellis:2007qk,Ossola_2007, Bern:1994cg}
\begin{align}
	I^{1-loop}=\sum_{i_{d_0+1}} C^{i_{d_0+1}} I_{d_0+1}^{i_{d_0+1}}+\sum_{i_{d_0}} C^{i_{d_0}} I_{d_0}^{i_{d_0}}+\cdots + \sum_{i_1} C^{i_1} I_1^{i_1}, \label{reduction_d0}
	\end{align}
where $i_s$ is the set of propagators appearing in the master integrals. The coefficient $C^{i_s}$ ($s=1,\cdots,d_0+1$) is simply a rational function of some Lorentz invariant such like the scalar product of external momenta while the terms $I_s^{i_s}$ are the $s$-gon scalar integral. With the general expansion \eref{reduction_d0}, the computation of general one-loop amplitudes  has been switched to determining those coefficients of $C^{i_s}$. Many tools have been invented to shovel the brambles, such as integration-by-parts (IBP) \cite{Chetyrkin:1981qh,Tkachov:1981wb}, PV reduction~\cite{Passarino:1978jh}, OPP reduction~\cite{Ossola:2006us,Ossola:2007bb,Ellis:2007br,Ossola_2007}, Unitarity cut \cite{Bern:1994zx,Bern:1994cg,Britto:2004nc,Britto:2005ha,Campbell:1996zw,Bern:1997sc,Denner:2005nn,Anastasiou:2006gt,Britto:2010um} etc.

All these methods can be divided into two categories, i.e., the reduction at the integrand level or the integral level. For reduction at the integrand level,
\cite{Ossola_2007} shows how to extract the coefficients of the 4-, 3-, 2- and
  1-point one-loop
scalar integrals from the full one-loop integrand of arbitrary scattering processes in an algebraical way. For the reduction at the integral level, an efficient way is the unitarity cut method. The main idea is to compare the imaginary part of two side of \eref{reduction_d0}. However, since the loop-integral is well-defined using the
dimensional regularization, the unitarity cut method in pure  4D need to  generalize  to $(4-2\eps)$-dimension, which has been done in \cite{Anastasiou:2006jv,Anastasiou:2006gt}. Based on this generalization, the analytic expressions for reduction coefficients (except the tadpole coefficients) have been derived in a series of  papers \cite{Britto:2006fc,Britto:2007tt,Britto:2008vq,Britto:2008sw,Feng:2013sa}.

In our previous work \cite{Feng:2021enk}, we reconsider the problem by introducing differential operators $\mathcal{D}$ and $\mathcal{T}$. We first introduce an auxiliary vector $R^\mu$ and reduce it to master integrals, then applying differential operators to the integrals with respect to $R$. By comparing two sides of the expansion, We will achieve the recursion relations of those coefficients of master integrals in differential form. With the knowledge of the algebraical structure of the reduction coefficients, we transform those differential equation form  relations into algebraical form. In~\cite{Feng:2021enk}, we solve the remaining unsolved tadpole coefficients by this method. In this paper, we will provide a general algorithm for calculating all reduction coefficients for a general tensor one-loop integral and give the explicit analytic results.

In section \ref{sec:recursion relation}, we will review the derivation of
the differential equations of reduction coefficients and show how to get the recursion relations of expansion
coefficients. In section  \ref{sec:expansion coefficients}, we will solve the recursion relations of reduction coefficients
in the general case. In section  \ref{sec:example}, we provide some examples and summarize the algorithm
for calculating the reduction of a general tensor one-loop Feynman integral. In appendix \ref{sec:More Example}, we list all reduction coefficients for tensor triangles, boxes and pentagons with rank $1$ and $2$.

\section{Differential equations and recursion relations}
\label{sec:recursion relation}
We will review the differential operators in \cite{Feng:2021enk} and show the way we obtain the recursion relations of every reduction coefficients. Starting with the following general one-loop $m$-rank tensor integral with $n+1$ propagators
\bea
I^{\mu_1\cdots \mu_m}_{n+1}
= \int {d^D\ell\over (2\pi)^D} \frac{\ell^{\mu_1}\ell^{\mu_2}\cdots \ell^{\mu_m}}{P_0P_1\cdots P_{n}},~\label{Idea-1-1-0}
\eea
where the $i$-th propagator is $P_i=(\ell-K_i)^2-M_i^2$ with setting $K_0=0$, we introduce an auxiliary vector $R^{\mu}$ and contract $\ell$ with $R$ on \eqref{Idea-1-1-0} $m$ times to get
\begin{align}
I^{(m)}_{n+1}[R]
\equiv& 2^m I^{\mu_1\cdots \mu_m}_{n+1} R_{\mu_1}\cdots R_{\mu_m}
=\int {d^D\ell\over (2\pi)^D} { (2\ell\cdot R)^m\over P_0P_1\cdots P_n}.~\label{Idea-1-1}
\end{align}
%
The vector $R$ lies in the same dimension as the $\ell$ does. The \eref{Idea-1-1} contains all information as in \eqref{Idea-1-1-0} but with much simpler organization of tensor structure. In this paper, we will focus on  $D=(4-2\epsilon)$-dimensional space, although our method can obviously be applied to arbitrary dimension. With this assumption, 
 the integral $I^{(m)}_{n+1}[R]$ is reduced to the linear combination of pentagon, box, triangle, bubble and tadpole master integrals
\begin{align}
I^{(m)}_{n+1}[R]
=& \sum_{a_1,a_2,\cdots,a_5} C^{(a_1,a_2,\cdots,a_5)}(m\vert n) I_5[a_1,a_2,\cdots,a_5] \notag\\
&+\sum_{a_1,a_2,a_3,a_4} C^{(a_1,a_2,a_3,a_4)}(m\vert n) I_4[a_1,a_2,a_3,a_4] +\cdots + \sum_{a_1} C^{(a_1)}(m\vert n) I_1[a_1].\label{reduction-formula}
\end{align}
The reduction coefficients $C^{(a_1,\cdots,a_r)}(m\vert n), 1\leq r\leq 5$ are the rational functions of external momenta $K_i$, masses $M_i$, and vector $R$. The summation in \eqref{reduction-formula} covers all possible combinations of $r$ propagators $\{P_{a_1},,,P_{a_r}\}\subseteq\{P_0,P_1,...,P_n\}$. We will use the abbreviation $C^{i_r}(m\vert n)$ instead of $C^{(a_1,...,a_r)}(m\vert n)$ with index set $i_r=\{a_1,...,a_r\}\subseteq\{0,1,2,...,n\}$.
It is easy to see vector $R$ only appears in the numerator of $C^{i_r}$ with the form $R\cdot R$ or $R\cdot K_i,\ i=1,2,...,n$. We introduce the following two operators
\bea
\mathcal{D}_i\equiv K_i\cdot {\d \over \d R},~~i=1,...,n;~~~~ \mathcal{T}\equiv \eta^{\mu\nu}{\d\over \d R^\mu}
{\d \over \d R^\nu}.\label{def-diffe}
 \eea
%
We take the derivative on the both sides of \eqref{reduction-formula}  by these two operators. The left-hand side will be
\begin{align}
\mathcal{D}_i I^{(m)}_{n+1}[R]
& =  m I^{(m-1)}_{n+1;\widehat{0}}- m I^{(m-1)}_{n+1;\widehat{i}}+ m f_i I^{(m-1)}_{n+1},  \notag\\
\mathcal{T} I^{(m)}_{n+1} [R]
& =  4m (m-1)M_0^2 I^{(m-2)}_{n+1}+ 4m(m-1)I^{(m-2)}_{n+1;\widehat{0} }, ~~~\label{operator_action}
\end{align}
where the constant $f_i\equiv M^2_0+K^2_i-M^2_i$, and

\begin{equation}
I^{(m-1)}_{n+1;\widehat{i}}[R]=\int {d^D\ell\over (2\pi)^D} { (2\ell\cdot R)^{m-1}\over P_0P_1\cdots P_{i-1}P_{i+1}\cdots P_n}
\end{equation}
i.e.,  the $i$-th propagator has been removed. 
For the right-hand side of \eqref{reduction-formula}, since the master integrals contains no $R$, the operators will directly act on coefficients $C^{i_r}(m\vert n)$. Therefore, We have the following equations
\begin{align}
	\sum_{s=1}^5 \sum_{i_s} \left(\mathcal{D}_{i}C^{i_s}(m\vert n) \right) I_s^{i_s}=&m \sum_{s=1}^5  \sum_{i_s}\left( C^{i_s}(m-1\vert n;\widehat{0})-C^{i_s}(m-1\vert n;\widehat{i})+f_iC^{i_s}(m-1\vert n)\right) I_{s}^{i_s},\label{Di_equation}
	\end{align}
and
\begin{align}
	\sum_{s=1}^5 \sum_{i_s} \left(\mathcal{T}C^{i_s}(m\vert n) \right) I_s^{i_s}
	=& 4m (m-1) \sum_{s=1}^5  \sum_{i_s} \left (C^{i_s}(m-2\vert n;\widehat{0})+M_0^2C^{i_s}(m-2\vert n)\right) I_{s}^{i_s},\label{T-equation}
\end{align}
%
where $C^{i_s}(m-1;\widehat{i})$ is the coefficient of the master integrals $I^{i_s}_{s+1}$ in the reduced expansion of tensor integral $I^{(m-1)}_{n+1;\widehat{i}}[R]$. Assuming that all the reduction coefficient of tensor integral $I^{(m')}_{n'+1}$ with either $m'<m$ or $n'<n$ are known already,  we can get a series of differential equations of $C^{i_r}(m\vert n)$ by comparing the coefficients of each master integral of both side of \eqref{Di_equation} and \eqref{T-equation}. Without loss of generality, we choose $i_r=(0,1,...,r)$\footnote{For simplicity, we will consider the reduction coefficients of master integrals with propagator $P_0$. Other cases can be obtained either by permutations or by momentum shifting. The details are shown in section \ref{sec:3.2}.
}. Then we have
\begin{align}
\mathcal{T} C^{(0,1,\cdots,r)}(m\vert n)= & 4m(m-1) M_0^2 C^{(0,1,\cdots,r)}(m-2\vert n), \label{differential_equation_T}
\end{align}
and
\begin{align}
\mathcal{D}_i C^{(0,1,\cdots,r)}(m\vert n)=& -m  C^{(0,1,\cdots,r)}(m-1\vert n;\widehat{i}) + mf_i C^{(0,1,\cdots,r)}(m-1\vert n).\label{differential_equation_D}
\end{align}
In the equation \eref{differential_equation_D}, $C^{(0,1,\cdots,r)}(m-1\vert n;\widehat{i})$ is the reduction coefficient
of the master integral $I_{r+1}^{(0,1,...,r)}$.  Since $\widehat{i}$ means the propagator $P_i$ has been removed, $C^{(0,1,\cdots,r)}(m-1\vert n;\widehat{i})$ is zero  when $i\leq r$. 

Similar to the idea used in  \cite{Feng:2021enk}, we do not solve
 the differential equations directly, but expand the reduction coefficients
 according to its tensor structure
\begin{align}
	C^{(0,1,\cdots,r)}(m\vert n)
	=& \sum_{2a_0+\sum_{k=1}^na_k=m}\Bigg\{ c^{(0,1,\cdots,r)}_{a_0,a_1,\cdots,a_n}(m) (M_0^2)^{a_0+r-n}\prod_{k=0}^n s_{0k}^{a_k}\Bigg\} ,\label{coefficient_expansion_1}
	\end{align}
where the notation $s_{00}\equiv(R\cdot R),s_{0i}\equiv(R\cdot K_i)$. The summation condition $2a_0+\sum_{k=1}^{n} a_k=m$ guarantees vector $R$ appears $m$ times. The exponent of $M^2_0$ makes  $c^{(0,1,\cdots,r)}_{a_0,a_1,\cdots,a_n}(m)$ dimensionless. The expansion coefficients $c^{(0,1,\cdots,r)}_{a_0,a_1,\cdots,a_n}(m)$\footnote{Note that we use capital $C$ to represent reduction coefficient while use lower case letter $c$ to represent expansion coefficient. } can only be a rational function of $(K_i\cdot K_j),i,j\neq 0$ and $M_i^2, (i=0,1,\cdots,n)$. Moreover, $c^{(0,1,\cdots,r)}_{a_0,a_1,\cdots,a_n}(m)$  vanish if any  $a_k<0,k=0,1,\cdots,n$. 

For $C^{(0,1,...,r)}(m-1\vert n;\widehat{i})$ in \eref{differential_equation_D}, the expansion is
\begin{align}
& C^{(0,1,\cdots,r)}(m-1\vert n;\widehat{i})\nn
=&\sum_{2a_0+\sum_{k=1,k\neq i}^n a_k=m-1}  c^{(0,1,\cdots,r)}_{a_0,\cdots,a_{i-1},a_{i+1},\cdots,a_n}(m-1;\widehat{i})(M_0^2)^{a_0+r-n}\prod_{k=0,k\neq i}^n s_{0k}^{a_k}\nn 
=&  \sum_{2a_0+\sum_{k=1}^n a_k=m-1}   \delta_{0a_i}~   c^{(0,1,\cdots,r)}_{a_0,\cdots,\widehat{a_i},\cdots,a_n}(m-1;\widehat{i})   ~  (M_0^2)^{a_0+r-n} \prod_{k=0}^n s_{0k}^{a_k}.\label{coefficient_expansion_2}
\end{align}
The absence of term $s_{0i}=(R\cdot K_i)$ is because propagator $P_i$ has been removed. In the last line of \eref{coefficient_expansion_2}, we add a factor $\delta_{0a_i}$ to make the expression simpler. The subscript $\widehat{a_i}$ means index $a_i$ is absent.
 
%
 

To get the algebraic recursion relation for expansion coefficients, we need rewrite 
$\mathcal{D}_i$ and $\mathcal{T}$ in terms of
 \begin{align}
 \mathcal{D}_i
 =&K_i^{\mu}{\d \over \d R^{\mu}}=2s_{0i}{\d \over \d s_{00}}+\sum_{j=1}^ns_{ij}{\d \over \d s_{0j}},   \notag\\
 \mathcal{T}
 =&2D{\d \over \d s_{00}}+4s_{00}{\d^2\over \d s_{00}^2}+4\sum_{i=1}^ns_{0i}{\d \over \d s_{0i}}{\d \over \d s_{00}}+\sum_{i=1}^n\sum_{j=1}^ns_{ij}{\d \over \d s_{0i}}{\d \over \d s_{0j}}. ~~~~\label{differential_operators}
 \end{align}
 With above explanation, putting \eref{coefficient_expansion_1} and \eref{coefficient_expansion_2} to \eref{differential_equation_T} and \eref{differential_equation_D}, comparing the expansion coefficients of $\prod_{k=0}^n s_{0k}^{a_k}$, the two types of differential equations  \eref{differential_equation_T} and \eref{differential_equation_D} become
\begin{align}
 	&(m+1-\sum_{l=1}^ni_l) c^{(0,1,\cdots,r)}_{a_1,\cdots,a_i-1,\cdots,a_n}(m)
 	+\sum_{l=1}^n(a_l+1)\beta_{il}c^{(0,1,\cdots,r)}_{a_1,\cdots,a_l+1,\cdots,a_n} (m) \notag\\
 	&= m\alpha_j c^{(0,1,\cdots,r)}_{a_1,\cdots,a_n}(m-1)- m  \delta_{0 a_i}  c^{(0,1,\cdots,r)}_{a_1,\cdots, \widehat{a_i},\cdots, a_n}(m-1;\widehat{i}) \label{recursion_relation_1}
\end{align}
for the ${\cal D}$-type and 
 \begin{align}
 4m(m-1)c^{(0,1,\cdots,r)}_{a_1,\cdots,a_n}(m-2)
 =& (m-\sum_{k=1}^n a_k)(D+m+\sum_{k=1}^n a_k-2)c^{(0,1,\cdots,r)}_{a_1,\cdots,a_n} (m) \notag\\
 &+\sum_{0<i<j}2(a_i+1)(a_j+1)\b_{ij}c^{(0,1,\cdots,r)}_{a_1,\cdots,a_i+1,\cdots,a_j+1,\cdots,a_n}(m) \notag\\
 &+\sum_{i=1}^n(a_i+1)(a_i+2)\b_{ii}c^{(0,1,\cdots,r)}_{a_1,\cdots,a_i+2,\cdots,a_n}(m) \label{recursion_relation_2}
 \end{align}
for the ${\cal T}$-type, where $\alpha_i\equiv f_i/M_0^2, \beta_{il}\equiv s_{il}/M_0^2$ for simplicity. Again we need to emphasize $c^{(0,1,\cdots,r)}_{a_1,\cdots, \widehat{a}_i,\cdots, i_n}(m-1;\widehat{i})=0$  in the case $i\leq r$ for the same reason as discussed before. 
In \eref{recursion_relation_1} and \eref{recursion_relation_2} we have ignored the subscript $a_0$ because it has been uniquely determined by the restrictive condition $2a_0+\sum_{k=1}^n a_n=m$ in \eref{coefficient_expansion_1}.


\section{Algorithm for recursion relations}
\label{sec:expansion coefficients}

The recurrence relations \eref{recursion_relation_1} and \eref{recursion_relation_2} are the key relations throughout the whole paper. In this section we show  how to solve expansion coefficients by these two relations systematically.
 
\subsection{Reduction coefficient of \texorpdfstring{$I_{r+1}[0,1,\cdots,r]$}{Lg}}

First, we choose the master basis contains propagator $\{P_0,P_1,\cdot,P_r\}$. We start by rewriting  $\mathcal{D}$-type  relations \eref{recursion_relation_1} in a compact form as
\begin{align}
\widetilde{\boldsymbol{G}}~\boldsymbol{T}~\boldsymbol{c}^{(0,1,\cdots,r)}(a_1,\cdots,a_n;m)= \boldsymbol{O}^{(0,1,\cdots,r)}(a_1,\cdots,a_n;m),~~~\label{ot-1}
\end{align}
where $\widetilde{\boldsymbol{G}}=[\beta_{ij}]$ is the $n\times n$ rescaled Gram matrix and  $\boldsymbol{T}$ is a diagonal matrix 
 \begin{align}
&\boldsymbol{T}= \text{diag}(a_1+1,a_2+1,\cdots,a_n+1).\end{align}
The  $\boldsymbol{c}^{(0,1,\cdots,r)}(a_1,\cdots,a_n;m),  \boldsymbol{O}^{(0,1,\cdots,r)}(a_1,\cdots,a_n;m)$ are two vectors defined as
 \begin{align}
 &[\boldsymbol{c}^{(0,1,\cdots,r)}(a_1,\cdots,a_n;m)]_i= c^{(0,1,\cdots,r)}_{a_1,a_2,\cdots,a_i+1,\cdots, a_n}(m),
 \end{align}
 and
 \begin{align}
 [\boldsymbol{O}^{(0,1,\cdots,r)}(a_1,\cdots,a_n;m)]_i
 = &m\alpha_i c^{(0,1,\cdots,r)}_{a_1,\cdots,a_n}(m-1)- m  \delta_{0 a_i}  c^{(0,1,\cdots,r)}_{a_1,\cdots, \widehat{a}_i,\cdots, a_n}(m-1;\widehat{i}) \notag\\
 &-(m+1-\sum_{l=1}^n a_l) c^{(0,1,\cdots,r)}_{a_1,\cdots,a_i-1,\cdots,a_n}(m).
 \end{align}
The definition of these two vectors are purposely for the recurrence construction. The vector $\boldsymbol{c}$ contains coefficients with 
rank $m$ and subscript with the summation $1+\sum_i a_i$, while 
the vector $\boldsymbol{O}$  contains coefficients 
of  three different patterns: (1) the first term with coefficients of rank $m-1$; (2)  the second one with coefficients of master integrals with one less propagator and lower rank $m-1$ rank; (3) the third one with coefficients of same rank $m$, but the summation $-1+\sum_i a_i$ of subscript. By induction assumption,  the first  two terms are considered   to be known already. Thus, by rewriting \eref{ot-1} as
\begin{align}
\boldsymbol{c}^{(0,1,\cdots,r)}(a_1,\cdots,a_n;m)=\boldsymbol{T}^{-1} \widetilde{\boldsymbol{G}}^{-1}  \boldsymbol{O}^{(0,1,\cdots,r)}(a_1,\cdots,a_n;m), \label{inverse_formula}
\end{align}
we have established the recurrence relations between expansion coefficients
with higher summation of subscript and those of same rank but with the summation of subscript reduced by two.

Iteratively using  \eref{inverse_formula}, we have left two kinds of unknown expansion coefficients
\bea
&c_{0,0,\cdots,0}^{(0,1,\cdots,r)}(m), &m=2k,\nn
&c^{(0,1,\cdots,r)}_{1,0,\cdots,0};c^{(0,1,\cdots,r)}_{0,1,\cdots,0};\cdots;c^{(0,1,\cdots,r)}_{0,0,\cdots,1},&m=2k+1,
\eea
depending on the parity of $m$.
For the odd case $m=2k+1$, we solve $c^{(0,1,\cdots,r)}_{1,0,\cdots,0}$'s by \eref{inverse_formula} again. To see it, setting $a_1=\cdots=a_n=0$ in \eref{inverse_formula}, the left hand side becomes

\bea
 (c^{(0,1,\cdots,r)}_{1,0,\cdots,0}(2k+1),c^{(0,1,\cdots,r)}_{0,1,\cdots,0}(2k+1),\cdots,c^{(0,1,\cdots,r)}_{0,0,\cdots,1}(2k+1))^T,
\eea
while the right-hand side is
\bea
\boldsymbol{T}^{-1} \widetilde{\boldsymbol{G}}^{-1}\boldsymbol{O}^{(0,1,\cdots,r)}(0,\cdots,0;2k+1),
\label{odd_reduce_even}
\eea
where
\bea
[\boldsymbol{O}^{(0,1,\cdots,r)}(0,\cdots,0;2k+1)]_i
=& m\alpha_i c^{(0,1,\cdots,r)}_{0,\cdots,0}(2k)- (2k+1)  c^{(0,1,\cdots,r)}_{0,\cdots, \widehat{a}_i,\cdots, 0}(2k;\widehat{i})
\eea
since the third term $c^{(0,1,\cdots,r)}_{0,\cdots,-1,\cdots,0}(2k+1)$ vanishes. Therefore, we reduced to the problem of 
solving  $c^{(0,1,\cdots,r)}_{0,\cdots,0}(2k)$.

Determining the value of $c_{0,0,\cdots,0}^{(0,1,\cdots,r)}(2k)$ requires $\mathcal{T}$-type recursion relations. For $m=2k$ and $a_1=a_2=\cdots=a_n=0$, $\mathcal{T}$-type recursion  relation becomes
\begin{align}
8k(2k-1)c^{(0,1,\cdots,r)}_{0,\cdots,0}(2k-2)
=& 2k(D+2k-2)c^{(0,1,\cdots,r)}_{0,\cdots,0}(2k) +\sum_{0<i<j<n}2\b_{ij} c^{(0,1,\cdots,r)}_{0,\cdots,1,\cdots,1,\cdots,0}(2k) \notag\\
&+\sum_{i=1}^n 2 \b_{ii}c^{(0,1,\cdots,r)}_{0,\cdots,2,\cdots,0}(2k).   \label{even_reduction}
\end{align}
In $c^{(0,1,\cdots,r)}_{0,\cdots,1,\cdots,1,\cdots,0}(2k)$, indices $1$ appear in the both $i$-th and $j$-th positions, while in $c^{(0,1,\cdots,r)}_{0,\cdots,2,\cdots,0}(2k)$ index $2$ appears in the $i$-th position. For $c^{(0,1,\cdots,r)}_{0,\cdots,2,\cdots,0}(2k)$ and
$c^{(0,1,\cdots,r)}_{0,\cdots,1,\cdots,1,\cdots,0}(2k)$ in \eref{even_reduction}, we use 
 \eref{inverse_formula} again to reach $c^{(0,1,\cdots,r)}_{0,\cdots,0}(2k)$. Then we establish the relation reduced rank $2k$ to $2k-2$ 
\begin{align}
c^{(0,1,\cdots,r)}_{\underbrace{0,\cdots,0}_{n\ times}}(2k)
={2k-1\over  D+2 k-n-2}\left[(4-\boldsymbol{\a}^T\widetilde{\boldsymbol{G}}^{-1}\boldsymbol{\a})c_{\underbrace{0,\cdots,0}_{n\ times}}^{(0,1,\cdots,r)}(2k-2)+\boldsymbol{\a}^T\widetilde{\boldsymbol{G}}^{-1}\boldsymbol{c}_{\underbrace{0,\cdots,0}_{n-1\ times}}^{(0,1,\cdots,r)}(2k-2)\right],\label{Relation-T}
\end{align}
where  $\boldsymbol{\a}$ is a vector define as 
\bea
\left(\alpha\right)^T&=&(\a_1,\a_2,\cdots,\a_n)^T=\left(\frac{f_1}{M_0^2},\frac{f_2}{M_0^2},\cdots,\frac{f_n}{M^2_0}\right)^T.
\eea 
In the second term of right-hand side of \eref{Relation-T}, $\boldsymbol{c}_{\underbrace{0,\cdots,0}_{n-1\ times}}^{(0,1,\cdots,r)}(m)$ is a vector defined as

\bea
\left(\boldsymbol{c}_{\underbrace{0,\cdots,0}_{n-1\ times}}^{(0,1,\cdots,r)}(m)\right)^T&=&\left(c_{\underbrace{0,\cdots,0}_{n-1\ times}}^{(0,1,\cdots,r)}(m;\widehat{1}),c_{\underbrace{0,\cdots,0}_{n-1\ times}}^{(0,1,\cdots,r)}(m;\widehat{2}),\cdots,c_{\underbrace{0,\cdots,0}_{n-1\ times}}^{(0,1,\cdots,r)}(m;\widehat{n})\right)^T\nn
&=&\left(0,0,\cdots,0,c_{\underbrace{0,\cdots,0}_{n-1\ times}}^{(0,1,\cdots,r)}(m;\widehat{r+1}),\cdots,c_{\underbrace{0,\cdots,0}_{n-1\ times}}^{(0,1,\cdots,r)}(m;\widehat{n})\right)^T.
\eea
The zero of first $r$ components has been explained under the Eq\eref{differential_equation_D}.
Equation \eref{Relation-T} reduced rank $m$ by two. Furthermore, we see a propagator is removed in the second term of Right-Hand-Side. Therefore, if we utilize the $\mathcal{T}$-type recursion relation repeatedly, we will end with one of the following two cases. (1) The rank $m$ is reduced to zero, which related to the reduction coefficient of a master integral. So it is either 1 or 0. (2) One of the propagator $P_i,i\le r$ has been removed. In this case the coefficients must be zero because the master integral will not appear in the reduction.

To make a long story short, for $d_0=4$, we summarize the whole reduction process below.
\begin{itemize}
	\item Step 1: For a tensor integral with more than $5$ propagators, we reduce it to 5-,4-,3-,2-,1-gon tensor integral.

	\item Step 2: For an arbitrary rank $m_0 $, we take each $m\le m_0$ with arrangement from small to large. If $m$ is even, we calculate the expansion coefficients $c_{a_1,\cdots,a_n}^{(0,1,\cdots,r)}(m)$ in the order $\sum_{i=1}^n a_i=0,2,4,\cdots,m$ by using \eref{inverse_formula} and \eref{even_reduction}. If $m$ is odd, we calculate in the order $\sum_{i=1}^n a_i=1,3,5,\cdots,m$ by  \eref{inverse_formula}.

	\item Step 3: We continue the Step 2 until $m=m_0$.
	
	\item Final step: Combining all expansion coefficients $c_{a_1,\cdots,a_n}^{(0,1,\cdots,r)}(m_0)$, we obtain  the reduction coefficient by \eref{coefficient_expansion_1}.
\end{itemize}

\subsection{Calculate  general \texorpdfstring{$C^{(j_0,j_1,\cdots,j_r)}(m\vert n)$ from $C^{(0,1,\cdots,r)}(m\vert n)$}{Lg}}
\label{sec:3.2}

In this subsection, we will show how to obtain the reduction coefficients of other MIs from the result of $C^{(0,1,\cdots,r)}(m\vert n)$. Let us begin with the case that the Master Integral contains propagators $P_0$. It is obvious that tensor integral $I_{n+1}^{(m)}$ is invariant under a permutation of labels $\{1,2,\cdots, n\}$. 
 Then the reduction coefficients $C^{(0,j_1\cdots,j_r)}(m\vert n)$ is simply  given by a proper replacement   $\sigma:\{M_{i},K_{i}\}\to\{M_{j_i},K_{j_i}\},(i=1,2,\cdots,n)$
\bea
C^{(0,j_1\cdots,j_r)}(m\vert n)=\sigma C^{(0,1,\cdots,r)}(m\vert n).\label{Other-Coeff-WithP0}
\eea

Now the remaining part is those  MIs without  $P_0$. Note that by a loop momenta shift $\ell \to \ell+K_{j_0}$ we have
\bea
\begin{aligned}
I_{n+1}^{(m)}[R]\to &\int {d^D\ell\over (2\pi)^D}{(2\ell\cdot R+2 K_{j_0}\cdot R)^m\over (\ell^2-M_{j_0}^2)\left[(\ell+K_{j_0})^2-M_0^2\right]\prod_{i=1,i\not=j_0}\left[(\ell-(K_i-K_{j_0})^2-M_i^2)\right]}\\
=&\sum_{k=0}^m \left(
\begin{array}{c}
	m \\
	k \\
\end{array}
\right)(2R\cdot K_{j_0})^{m-k}\times \\
&\int {d^D\ell\over (2\pi)^D}{(2\ell\cdot R)^k\over (\ell^2-M_{j_0}^2)\left[(\ell+K_{j_0})^2-M_0^2\right]\prod_{i=1,i\not=j_0}\left[(\ell-(K_i-K_{j_0})^2-M_i^2)\right]}.
\end{aligned}
\eea
By variable substitution $K_{j_0}\to -K_{j_0},K_i\to K_i-K_{j_0},M_{j_0}\leftrightarrow M_0$ inside the integrand\footnote{Note that we don't substitute $K_{j_0}$ in $(2R\cdot K_{j_0})^{m-k}$.}, we arrive the same form as \eref{Idea-1-1-0} . Then we have
\bea
C^{(j_0,j_1,\cdots,j_r)}(m\vert n)&=&\sum_{k=0}^m \left(
\begin{array}{c}
	m \\
	k \\
\end{array}
\right)(2R\cdot K_{j_0})^{m-k}\left[\left.C^{(0,j_1,\cdots,j_r)}(k\vert n)\right\vert_{K_{j_0}\to -K_{j_0},K_i\to K_i-K_{j_0},M_{j_0}\leftrightarrow M_0}\right]\nn
&=&\left.\left[\sum_{k=0}^m \left(
\begin{array}{c}
	m \\
	k \\
\end{array}
\right)(-2R\cdot K_{j_0})^{m-k}C^{(0,j_1,\cdots,j_r)}(k\vert n)\right]\right\vert_{K_{j_0}\to -K_{j_0},K_i\to K_i-K_{j_0},M_{j_0}\leftrightarrow M_0}.\label{General-Coeff}\nn
\eea

\section{Examples}
\label{sec:example}

Having presented the general algorithm, in this section we will use various examples to demonstrate the use of the algorithm. In the first subsection, we will show how to reduce
any tensor bubble to the basis of scalar bubble and two scalar 
tadpoles. The reduction of  tensor triangles, tensor 
boxes and tensor pentagons of rank $1$ and $2$ has been given
in the Appendix. In the second subsection, we will show how to get the 
reduction coefficients of tensor box with rank $1$ to scalar triangles without $P_0$ from the result of $ C^{(0,1,2)}(1|3) $ by \eref{General-Coeff}.

\subsection{The reduction of tensor bubble}
The reduction of tensor bubble $I_{2}^{(m)}$ will contain three  MIs as below
\bea
Tadpoles:&&I_1[0],I_1[1],\nn
Bubbles:&&I_2[0,1],
\eea
and we have the expansion
\bea
I_{2}^{(m)}=C^{(0)}(m\vert 1)I_1[0]+C^{(1)}(m\vert 1)I_1[1]+C^{(0,1)}(m\vert 1)I_2[0,1].
\eea
The way to achieve $C^{(0)}(m\vert 1)$ have been given in~\cite{Feng:2021enk}.  Here we only provide how to calculate $C^{(1)}(m\vert 1)$ and $C^{(0,1)}(m\vert 1)$. The coefficient of $I_1[1]$ can be obtained by \eref{General-Coeff} from $C^{(0)}(m\vert 1)$. While for $I_2[0,1]$, there is only one subscript in the expansion coefficients. So the expansion of $C^{(0,1)}(m\vert 1)$ is
\bea
C^{(0,1)}(m\vert 1)=\sum_{i}c_i^{(0,1)}(m)[M_0^2(R\cdot R)]^{{m-i\over 2}}(R\cdot K_1)^i.\label{bubble expansion}
\eea
We have the corresponding $\mathcal{D}$-type recursion relation
\bea
c^{(0,1)}_{i+2}(m)&=&{1\over (i+2)\beta_{11}}\left(m\a_1c^{(0,1)}_{i+1}(m-1)-m\delta_{0,i+1}c^{(0,1)}(m-1)-(m-i)c^{(0,1)}_i(m)\right)\nn
&=&{1\over (i+2)\beta_{11}}\left(m\a_1c^{(0,1)}_{i+1}(m-1)-(m-i)c^{(0,1)}_i(m)\right),~\label{Bub-ToNextTerm}
\eea
and $\mathcal{T}$-type recursion relation

\bea
c_0^{(0,1)}(2r)
&=&\frac{2r-1}{2r+D-3}\left[ \left(4-\frac{\a_1^2}{\b_{11}} \right )c_0^{(0,1)}(2r-2) + \frac{\a_1}{\b_{11}} c^{(0,1)}(2r-2) \right]\nn
&=&\frac{2r-1}{2r+D-3} \left(4-\frac{\a_1^2}{\b_{11}} \right )c_0^{(0,1)}(2r-2),\label{Bub-T}
\eea
where $c^{(0,1)}(m)$ and $c^{(0,1)}(2r-2)$ without subscript stands for  $c^{(0,1)}_{\widehat{a_1}}(m)$ and $c^{(0,1)}_{\widehat{a_1}}(2r-2)$. These two terms vanish, because they come from the reduction coefficient of bubble $I_2[0,1]$ for a tensor tadpole, i.e., the propagator $P_1$ has been removed.

Now we show the result for  rank $m\leq 4$. The rank $m=0$ is trivial. For other ranks:
\begin{itemize}
	\item $m=1$
	
	\textbf{The reduction coefficients of tadpoles: $I_1[0],I_1[1]$}
	
	Using the result in~\cite{Feng:2021enk}, we have
	\bea C^{(0)}(1\vert 1)=-{R\cdot K_1\over  K_1^2}.
	\eea
	For $I_1[1]$, by choosing $j_0=1$ in \eref{General-Coeff},  we have
	\bea
	C^{(1)}(1\vert    1)&=&\left.C^{(0)}(1\vert 1)\right\vert_{K_1\to -K_1,M_0\leftrightarrow M_1}\nn
	&=&{R\cdot K_1\over K_1^2}.
	\eea
	\textbf{The reduction coefficients of  bubble: $I_2[0,1]$}
	
	 The expansion of $C^{(0,1)}(1\vert 1)$  becomes
	\bea
	C^{(0,1)}(1\vert 1)=c_1^{(0,1)}(1)(R\cdot K_1).
	\eea
    By \eref{Bub-ToNextTerm}, we have
	\bea
	c_1^{(0,1)}(1)&=&{1\over \beta_{11}}\left(\a_1c^{(0,1)}_{0}(0)-2c^{(0,1)}_{-1}(1)\right)\nn
	&=&{f_1\over s_{11}},
	\eea
	where the boundary conditions are $c^{(0,1)}_{0}(0)=1,c^{(0,1)}_{-1}(1)=0$. Then
	\bea
	C^{(0,1)}(1\vert 1)=c_1^{(0,1)}(1)R\cdot K_1=\frac{\left(K_1\cdot K_1+M_0^2-M_1^2\right) R\cdot K_1}{ K_1\cdot K_1}.
	\eea
	\item $m=2$
	
	\textbf{The reduction coefficients of tadpoles: $I_1[0],I_1[1]$}
	
	The reduction coefficient of tadpole $I_1[0]$ is 
	\bea
	C^{(0)}(2\vert 1)=\frac{\left(K_1\cdot K_1+M_0^2-M_1^2\right) \left(K_1\cdot K_1 R\cdot R-D \left(R\cdot K_1\right)^2\right)}{(D-1) \left(K_1\cdot K_1\right)^2}.
	\eea
	For $I_1[1]$, by choosing $j_0=1$ in \eref{General-Coeff},  we have
	\bea
	C^{(1)}(2\vert 1)&=&2(2R\cdot K_1)\left [\left.C^{(0)}(1\vert 1)\right\vert_{K_1\to -K_1,M_0\leftrightarrow M_1}\right]+\left.C^{(0)}(2\vert 1)\right\vert_{K_1\to -K_1,M_0\leftrightarrow M_1}\nn
	&=&{4 (R\cdot K_1)^2\over K_1^2}+\frac{\left(K_1\cdot K_1+M_1^2-M_0^2\right) \left(K_1\cdot K_1 R\cdot R-D \left(R\cdot K_1\right)^2\right)}{(D-1) \left(K_1\cdot K_1\right)^2}.\nn
	\eea
	\textbf{The reduction coefficients of  bubble: $I_2[0,1]$}
	
	The expansion of $C^{(0,1)}(2\vert 1)$  is
	\bea
	C^{(0,1)}(2\vert 1)=c_0^{(0,1)}(2)M_0^2s_{00}+c_2^{(0,1)}(2)s_{01}^2.
	\eea
	By setting $r=1$ in \eref{Bub-T} , we have
	\bea
	c_0^{(0,1)}(2)
	&=&\frac{1}{D-1} \left(4-\frac{\a_1^2}{\b_{11}} \right )c_0^{(0,1)}(0) \nn
	&=&{1\over D-1}\left(4-\frac{\a_1^2}{\b_{11}} \right )\nn
	&=&\frac{4}{D-1}-\frac{ f_1^2}{(D-1) M_0^2 s_{11}},
	\eea
	where the boundary conditions is $c^{(0,1)}_{0}(0)=1$.
	 
	By setting $i=0$ in \eref{Bub-ToNextTerm}, we have
	\bea
	c^{(0,1)}_{2}(2)&=&{1\over 2\beta_{11}}\left(2\a_1c^{(0,1)}_{1}(1)-2c^{(0,1)}_0(2)\right)\nn
	&=&{1\over \beta_{11}}\left(\alpha_1{f_1\over s_{11}}-{1\over D-1}\left(4-\frac{\a_1^2}{\b_{11}} \right )\right)\nn
	&=&\frac{D f_1^2}{(D-1) s_{11}^2}-\frac{4 M_0^2}{(D-1) s_{11}},
	\eea
	where $c^{(0,1)}_1(1)$ has been presented in the case $m=1$. Then
	\bea
	C^{(0,1)}(2\vert 1)
	&=&\frac{s_{01}^2 \left(D f_1^2-4 M_0^2 s_{11}\right)+s_{00} s_{11} \left(4 M_0^2 s_{11}-f_1^2\right)}{(D-1) s_{11}^2}.
	\eea
	\item $m=3$
	
	\textbf{The reduction coefficients of tadpoles: $I_1[0],I_1[1]$}

	The reduction coefficient of tadpole $I_1[0]$ is 
	\bea
	C^{(0)}(3\vert 1)=\frac{f_1^2 \left(3 s_{00} s_{01} s_{11}-(D+2) s_{01}^3\right)}{(D-1) s_{11}^3}+\frac{4 M_0^2 s_{01} \left(2 s_{01}^2-3 s_{00} s_{11}\right)}{D s_{11}^2}.
	\eea
	By choosing $j_0=1$ in \eref{General-Coeff},  we have
	\bea
	C^{(1)}(3\vert 1)&=&3(2 R\cdot K_1)^2\left [\left.C^{(0)}(1\vert 1)\right\vert_{K_1\to -K_1,M_0\leftrightarrow M_1}\right]+3(2 R\cdot K_1)\left [\left.C^{(0)}(2\vert 1)\right\vert_{K_1\to -K_1,M_0\leftrightarrow M_1}\right]\nn
	&&+C^{(0)}(3\vert 1)\left [\left.C^{(0)}(3\vert 1)\right\vert_{K_1\to -K_1,M_0\leftrightarrow M_1}\right]\nn
	&=&\frac{s_{01} \left(7 D^2 s_{01}^2+12 D M_1^2 s_{00}-10 D s_{01}^2-12 M_1^2 s_{00}\right)}{(D-1) D s_{11}}+\frac{(D+2) \left(M_0^2-M_1^2\right)^2 s_{01}^3}{(D-1) s_{11}^3}\nn
	&&+\frac{4 \left(D M_0^2-D M_1^2-2 M_1^2\right) s_{01}^3}{Ds_{11}^2}-\frac{3 \left(M_0^2-M_1^2\right)^2 s_{00} s_{01}}{(D-1)s_{11}^2}+\frac{3 s_{00} s_{01}}{D-1}.
	\eea
	\textbf{The reduction coefficients of  bubble: $I_2[0,1]$}
	
	The expansion of $C^{(0,1)}(3\vert 1)$  is
	\bea
	C^{(0,1)}(3\vert 1)=c_1^{(0,1)}(3)M_0^2s_{00}s_{01}+c_3^{(0,1)}(3)s_{01}^3.
	\eea
	By setting $i=-1$ in \eref{Bub-ToNextTerm}, we have
	\bea
	c_1^{(0,1)}(3)&=&{1\over \beta_{11}}\left(3\a_1c^{(0,1)}_{0}(2)-4c_{-1}^{(0,1)}(3)\right)\nn
	&=&{1\over \beta_{11}}\left [3\a_1\left (\frac{4}{D-1}-\frac{ f_1^2}{(D-1) M_0^2 s_{11}}\right)\right]\nn
	&=&\frac{12 f_1}{(D-1) s_{11}}-\frac{3 f_1^3}{(D-1) M_0^2 s_{11}^2},
	\eea
	where we have used $c_{-1}^{(0,1)}(3)=0$ and the result of expansion coefficient $c^{(0,1)}_{0}(2)$ in the case $m=2$. 
	
	By setting $i=1$ in \eref{Bub-ToNextTerm}, we have
	\bea
	c^{(0,1)}_{3}(3)&=&{1\over 3\beta_{11}}\left(3\a_1c^{(0,1)}_{2}(2)-2c^{(0,1)}_1(3)\right)\nn
	&=&{1\over 3\b_{11}}\left[3\a_1\left (\frac{D f_1^2}{(D-1) s_{11}^2}-\frac{4 M_0^2}{(D-1) s_{11}}\right)-2\left (\frac{12 f_1}{(D-1) s_{11}}-\frac{3 f_1^3}{(D-1) M_0^2 s_{11}^2}\right)\right]\nn
	&=&\frac{(D+2) f_1^3}{(D-1) s_{11}^3}-\frac{12 f_1 M_0^2}{(D-1) s_{11}^2}.
	\eea
  Then the reduction coefficient is
	\bea
	C^{(0,1)}(3\vert 1)&=&\frac{f_1 \left(s_{01}^3 \left((D+2) f_1^2-12 M_0^2 s_{11}\right)+3 s_{00} s_{11} s_{01} \left(4 M_0^2 s_{11}-f_1^2\right)\right)}{(D-1) s_{11}^3}.
	\eea
	\item $m=4$
	
	\textbf{The reduction coefficients of tadpoles: $I_1[0],I_1[1]$}

	The reduction coefficient of tadpole $I_1[0]$ is 
	\bea
	C^{(0)}(4\vert 1)&=&-\frac{3 f_1 s_{00} \left(8 D^2 M_0^2 s_{01}^2+D f_1^2 s_{00}+16 D M_0^2 s_{01}^2-16 M_0^2 s_{01}^2\right)}{(D-1) D (D+1) s_{11}^2}\nn
	&&+\frac{2 (D+2) f_1 s_{01}^2 \left(3 D f_1^2 s_{00}+10 D M_0^2 s_{01}^2-8 M_0^2 s_{01}^2\right)}{(D-1) D (D+1) s_{11}^3}+\frac{12 (2 D-1) f_1 M_0^2 s_{00}^2}{(D-1) D (D+1) s_{11}}\nn
	&&-\frac{(D+2) (D+4) f_1^3 s_{01}^4}{(D-1) (D+1) s_{11}^4}.
	\eea
	By choosing $j_0=1$ in \eref{General-Coeff},  we have
	\bea
	C^{(1)}(4\vert 1)
	&=&4(2 R\cdot K_1)^3\left [\left.C^{(0)}(1\vert 1)\right\vert_{K_1\to -K_1,M_0\leftrightarrow M_1}\right]+6(2 R\cdot K_1)^2\left [\left.C^{(0)}(2\vert 1)\right\vert_{K_1\to -K_1,M_0\leftrightarrow M_1}\right]\nn
	&&+4(2R\cdot K_1)\left [\left.C^{(0)}(3\vert 1)\right\vert_{K_1\to -K_1,M_0\leftrightarrow M_1}\right]+\left [\left.C^{(0)}(4\vert 1)\right\vert_{K_1\to -K_1,M_0\leftrightarrow M_1}\right]\nn
	&=& \widetilde{f}_1 \left(\frac{4 \left(5 D^2+6 D-8\right) M_1^2 s_{01}^4}{D \left(D^2-1\right) s_{11}^3}+\frac{12 s_{00} \left((2 D-1) M_1^2 s_{00}+2 D (D+1) s_{01}^2\right)}{D \left(D^2-1\right) s_{11}}\right)\nn
	&&-\widetilde{f}_1\frac{24 \left(\left(D^2+2 D-2\right) M_1^2 s_{00} s_{01}^2+D^2 (D+1) s_{01}^4\right)}{D \left(D^2-1\right) s_{11}^2}\nn
	&&+ \widetilde{f}_1^3 \left(-\frac{\left(D^2+6 D+8\right) s_{01}^4}{\left(D^2-1\right) s_{11}^4}+\frac{6 (D+2) s_{00} s_{01}^2}{\left(D^2-1\right) s_{11}^3}-\frac{3 s_{00}^2}{\left(D^2-1\right) s_{11}^2}\right)\nn
	&& +\widetilde{f}_1^2 \left(\frac{8 (D+2) s_{01}^4}{(D-1) s_{11}^3}-\frac{24 s_{00} s_{01}^2}{(D-1) s_{11}^2}\right)-\frac{64 M_1^2 s_{01}^4}{D s_{11}^2}+\frac{32 s_{01}^2 \left(\frac{3 M_1^2 s_{00}}{D}+s_{01}^2\right)}{s_{11}},
	\eea
	where $\widetilde{f}_1=K_1^2+M_1^2-M_0^2$.
	
	\textbf{The reduction coefficients of  bubble: $I_2[0,1]$}
	
	The expansion of $C^{(0,1)}(4\vert 1)$  is
	\bea
	C^{(0,1)}(4\vert 1)=c_0^{(0,1)}(4)M_0^4s_{00}^2+c_2^{(0,1)}(4)M_0^2s_{00}s_{01}^2+c_4^{(0,1)}(4)s_{01}^4.
	\eea
	By setting $r=2$ in \eref{Bub-T}, we have
	\bea
	c_0^{(0,1)}(4)&=&\frac{4}{D+1} \left(4-\frac{\a_1^2}{\b_{11}} \right )c_0^{(0,1)}(2)\nn
	&=&{4\over D+1}\left[\left(4-\frac{\a_1^2}{\b_{11}} \right )\left(\frac{4}{D-1}-\frac{ f_1^2}{(D-1) M_0^2 s_{11}}\right)\right]\nn
	&=&\frac{3 f_1^4}{\left(D^2-1\right) M_0^4 s_{11}^2}-\frac{24 f_1^2}{\left(D^2-1\right) M_0^2 s_{11}}+\frac{48}{D^2-1},
	\eea
    By setting $i=0,2$ in	\eref{Bub-ToNextTerm}, we calculate $c_2^{(0,1)}(4),c_4^{(0,1)}(4)$ iteratively
	\bea
	c^{(0,1)}_{2}(4)&=&{1\over 2\beta_{11}}\left(4\a_1c^{(0,1)}_{1}(3)-4c^{(0,1)}_0(4)\right)\nn
	&=&{1\over 2\beta_{11}}\Bigg[4\a_1\left(\frac{(D+2) f_1^3}{(D-1) s_{11}^3}-\frac{12 f_1 M_0^2}{(D-1) s_{11}^2}\right)\nn
	&&-4\left(\frac{3 f_1^4}{\left(D^2-1\right) M_0^4 s_{11}^2}-\frac{24 f_1^2}{\left(D^2-1\right) M_0^2 s_{11}}+\frac{48}{D^2-1}\right)\Bigg]\nn
	&=&-\frac{6 (D+2) f_1^4}{\left(D^2-1\right) M_0^2 s_{11}^3}+\frac{24 (D+3) f_1^2}{\left(D^2-1\right) s_{11}^2}-\frac{96 M_0^2}{\left(D^2-1\right) s_{11}},
	\eea
	\bea
	c_4^{(0,1)}(4)&=&{1\over 4\beta_{11}}\left(4\a_1c^{(0,1)}_{3}(3)-2c^{(0,1)}_2(4)\right)\nn
	&=&{1\over 4\beta_{11}}\Bigg [4\a_1\left(\frac{(D+2) f_1^3}{(D-1) s_{11}^3}-\frac{12 f_1 M_0^2}{(D-1) s_{11}^2}\right)\nn
	&&-2\left(-\frac{6 (D+2) f_1^4}{\left(D^2-1\right) M_0^2 s_{11}^3}+\frac{24 (D+3) f_1^2}{\left(D^2-1\right) s_{11}^2}-\frac{96 M_0^2}{\left(D^2-1\right) s_{11}}\right)\Bigg]\nn
	&=&-\frac{24 (D+2) f_1^2 M_0^2}{\left(D^2-1\right) s_{11}^3}+\frac{\left(D^2+6 D+8\right) f_1^4}{\left(D^2-1\right) s_{11}^4}+\frac{48 M_0^4}{\left(D^2-1\right) s_{11}^2}.
	\eea
    Where we have used the results of the expansion coefficients with lower rank. Then the reduction coefficient is
	\bea
	C^{(0,1)}(4\vert 1)&=&-\frac{24 f_1^2 M_0^2 \left(s_{01}^2-s_{00} s_{11}\right) \left((D+2) s_{01}^2-s_{00} s_{11}\right)}{\left(D^2-1\right) s_{11}^3}+\frac{48 M_0^4 \left(s_{01}^2-s_{00} s_{11}\right)^2}{\left(D^2-1\right) s_{11}^2}\nn
	&&+\frac{f_1^4 \left(\left(D^2+6 D+8\right) s_{01}^4-6 (D+2) s_{00} s_{11} s_{01}^2+3 s_{00}^2 s_{11}^2\right)}{\left(D^2-1\right) s_{11}^4}.
	\eea
\end{itemize}
\subsection{Reduce tensor box to scalar triangles}
We will consider the reduction coefficients of triangle MIs of tensor integral $I^{(1)}_4$ as another example to illustrate the algorithm in section \ref{sec:3.2}. For simplicity, we denote $G(i_1,i_2,\cdots,i_s;j_1,j_2,\cdots,j_r)$ as the determinant of the Gram matrix $G$ with entries $G_{ab}=K_a\cdot K_b=s_{ab}$. Specially, we denote $ G(i_1,i_2,\cdots,i_s)\equiv G(i_1,i_2,\cdots,i_s;i_1,i_2,\cdots,i_s)$.

The reduction coefficient of the scalar triangle $I_3[0,1,2]$ is
\bea
C^{(0,1,2)}(1\vert 3)=-\frac{G(2,3;1,2) s_{01}-G(1,3;1,2) s_{02}+G(1,2;1,2) s_{03}}{ G(1,2,3)}.
\eea
The  reduction coefficients of $I_3[0,1,3],I_3[0,2,3]$ are easy to obtained by simply changing labels $\{1,2,3\}\rightarrow \{1,3,2\}$ and $\{1,2,3\}\rightarrow \{2,1,3\}$ respectively:
\bea
C^{(0,1,3)}(1\vert 3)&=&\left.C^{(0,1,2)}(1\vert 3)\right\vert_{K_2\leftrightarrow K_3, M_2\leftrightarrow M_3}\nn
&=&-\frac{G(3,2;1,3) s_{01}-G(1,2;1,3) s_{03}+G(1,3;1,3) s_{02}}{ G(1,2,3)},\nn
C^{(0,2,3)}(1\vert 3)&=&\left.C^{(0,1,2)}(1\vert 3)\right\vert_{K_1\leftrightarrow K_2, M_1\leftrightarrow M_2}\nn
&=&-\frac{G(2,1;3,2) s_{03}-G(3,1;3,2) s_{02}+G(3,2;3,2) s_{01}}{ G(1,2,3)}.
\eea
Now we consider the reduction coefficient of the triangle without $P_0$, i.e.,  $I_3[1,2,3]$. In \eref{General-Coeff}, choosing $j_0=3,j_1=1,j_2=2$, we have
\bea
C^{(1,2,3)}(1\vert 3)
&=&C^{(3,1,2)}(1\vert 3)=(2R\cdot K_3)\left.C^{(0,1,2)}(0\vert 3)\right\vert_{K_1\to K_1-K_3,K_2\to K_2-K_3,K_3\to -K_3,M_0\leftrightarrow M_3}\nn
&&+\left.C^{(0,1,2)}(1\vert 3)\right\vert_{K_1\to K_1-K_3,K_2\to K_2-K_3,K_3\to -K_3,M_0\leftrightarrow M_3}\nn
&=&\frac{G(K_2-K_3,K_3;K_1-K_3,K_2-K_3) (s_{01}-s_{03})}{ G(K_1-K_3,K_2-K_3,K_3;K_1-K_3,K_2-K_3,K_3)}\nn
&&+\frac{-G(K_1-K_3,K_3;K_1-K_3,K_2-K_3) (s_{02}-s_{03})}{ G(K_1-K_3,K_2-K_3,K_3;K_1-K_3,K_2-K_3,K_3)}\nn
&&+\frac{G(K_1-K_3,K_2-K_3;K_1-K_3,K_2-K_3) s_{03}}{ G(K_1-K_3,K_2-K_3,K_3;K_1-K_3,K_2-K_3,K_3)},
\eea
where  $C^{(0,1,2)}(0\vert 3)=0$ is the reduction coefficient of a triangle MI from a box MI.
\section{Discussion}
In this paper, we show how to use the differential operators to get the analytical expressions for the reduction coefficients of all master basis. By these operators, one can establish the recurrence relations about reduction coefficients in differential equation form. Another crucial step in this method is that we use the information of tensor structure to avoid solving the intricate differential equations directly. 

As we have reviewed, in \cite{Britto:2006fc,Britto:2007tt,Britto:2008vq}, the analytical expressions for reduction coefficients can be solved by unitary method. However, there are some differences between these two approaches. 
\begin{itemize}
	\item The first difference is that  the expression given by unitarity cut method is written using the spinor formalism, while results in this paper use the traditional Lorentz invariant contractions.
	
	\item The second difference is that in unitarity cut method, we have assumed the
	external momenta to be in pure $4D$ and only loop momentum is in general $(4-2\eps)$ dimension. For our new method, there is no such a constraint and the external momenta can be in $4D$ or in $(4-2\eps)$ dimension. 
	
	\item The third difference is that results in the paper are defined in an iterated way, while expressions given by unitarity cut method are just one equation (although the differentiation has the spirit of iteration).
	\item The fourth difference is that expressions by unitarity cut method using the input of arbitrary forms, while the one in this paper using the standard input given in \eref{Idea-1-1}.  The difference has a potential and huge impact on computation efficiency. The reason is that with the development of on-shell program, it is well known that tree-level amplitudes will be significantly simplified if we use spinor variables with spurious poles, such as these given by the recursion relation \cite{Britto:2004ap,Britto:2005fq}. Thus, it will be desirable to cooperate these advantages of unitarity cut method to our current new strategy.
\end{itemize}

Before ending this paper, let us emphasize that the purpose of the
paper is to establish an independent and complete reduction framework for one-loop integrals using auxiliary vector $R$. In the previous work
\cite{Feng:2021enk},
we have discussed the reduction coefficients of tadpoles. In this paper we have complete the coefficients of other basis. However,
for these two works, we have assumed the power of each propagator
is just one. To be a complete reduction framework, we need to
give the reduction of integrals with arbitrary tensor 
structures and propagators having arbitrary 
powers. We will show in the upcoming paper how to achieve this. 
After completing the framework of reduction with auxiliary vector $R$, we can discuss various limit cases, like the massless limit or the vanishing of Gram determinant, which will be presented in another paper soon. Another direction is to apply our new framework to higher loops. But unlike one-loop case, relations established by differentiation over $R$ are usually not enough. Besides, the master integrals are more complicated in higher-loop integrals. How to solve these
 difficulties will be another future project. 

\section*{Acknowledgments}
We are very grateful for Bo Feng's collaboration in the previous related paper and inspiring discussion. This work is supported by Qiu-Shi Funding and Chinese NSF funding under Grant No.11935013, No.11947301, No.12047502 (Peng Huanwu Center).

\appendix
\section{More examples}
\label{sec:More Example}
In this appendix, we provide more examples to illustrate our method.\footnote{All results have been checked with Fire6.~\cite{Smirnov:2008iw,Smirnov:2014hma,Smirnov:2013dia,Smirnov:2019qkx,Lee:2013mka}} There are three points we need to emphasize ahead.
\begin{itemize}
	\item For the tensor integral $I_{n+1}^{(m)}$,  we only list the reduction coefficient $C^{i_r}$ for $m\ge n-\abs{i_r}$, because there are not enough  $\ell \cdot R$ in the nominator to cancel $n-\abs{i_r}$ propagators for $m<n-\abs{i_r}$.

    \item We merely list the results of $C^{(0,1,\cdots,r)}(m\vert n),C^{(1,2,\cdots,r+1)}(m\vert n),0\le r\le 4$ due to the permutation symmetry:
\begin{align}
Tadpoles: C^{(i)}(m\vert n)&=\left.C^{(1)}(m\vert n)\right\vert_{1\leftrightarrow i}\nn
Bubbles: C^{(0,i)}(m\vert n)&=\left.C^{(0,1)}(m\vert n)\right\vert_{1\leftrightarrow i}\nn
C^{(i,j)}(m\vert n)&=\left.C^{(1,2)}(m\vert n)\right\vert_{1\leftrightarrow i,2\leftrightarrow j}\nn
Triangles: C^{(0,i,j)}(m\vert n)&=\left.C^{(0,1,2)}(m\vert n)\right\vert_{1\leftrightarrow i,2\leftrightarrow j}\nn
C^{(i,j,k)}(m\vert n)&=\left.C^{(1,2,3)}(m\vert n)\right\vert_{1\leftrightarrow i,2\leftrightarrow j,3\leftrightarrow k}\nn
Boxes: C^{(0,i,j,k)}(m\vert n)&=\left.C^{(0,1,2,3)}(m\vert n)\right\vert_{1\leftrightarrow i,2\leftrightarrow j,3\leftrightarrow k}\nn
C^{(i,j,k,l)}(m\vert n)&=\left.C^{(1,2,3,4)}(m\vert n)\right\vert_{1\leftrightarrow i,2\leftrightarrow j,3\leftrightarrow k,4\leftrightarrow l}\nn
&\cdots
\end{align}
where $0<i<j<k<l$.

   \item  There is a  permutation symmetry about the expansion coefficient $c^{(j_0,\cdots,j_r)}_{a_1,\cdots,a_n}(m)$.  If  the Master Integral $I_{r+1}[j_0,j_1,\cdots,j_r]$ is invariant under a label permutation   $\sigma : \{1,2,\cdots,n\} \to \{\sigma(1),\sigma(2),\cdots,\sigma(n)\}$,  we have $c^{(j_0,\cdots,j_r)}_{i_1,\cdots,i_n}(m)=\sigma c^{(j_0,\cdots,j_r)}_{i_{\sigma^{-1}(1)},\cdots,i_{\sigma^{-1}(n)}}(m) $.

   For example, $I_4[0,1,2,3]$ is invariant under the label permutation  $\sigma:\{1,2,3,4\}\to \{3,1,2,4\}$, then we have
   \bea
   c_{1,2,4,5}^{(0,1,2,3)}(14)&=&\left.c_{2,4,1,5}^{(0,1,2,3)}(14)\right\vert_{\{1,2,3,4\}\to \{3,1,2,4\}}.
   \eea
\end{itemize}

\subsection{All reduction coefficients of tensor triangle with rank $m=1,2$}
The MI of a tensor triangle $I_3^{(m)}$ are
\bea
Tadpoles:&&I_1[0],I_1[1],I_1[2],\nn
Bubbles:&&I_2[0,1],I_2[0,2],I_2[1,2],\nn
Triangles:&&I_3[0,1,2].
\eea
Then the reduction of the tensor triangle is
\bea
I_{3}^{(m)}&=&C^{(0)}(m\vert 2)I_1[0]+C^{(1)}(m\vert 2)I_1[1]+C^{(2)}(m\vert 2)I_1[2]+C^{(0,1)}(m\vert 2)I_2[0,1]\nn
&&+C^{(0,2)}I_2[0,2]
+C^{(1,2)}I_2[1,2]+C^{(0,1,2)}I_3[0,1,2].
\eea
\begin{itemize}
	\item $m=1$
	
	\textbf{Reduction coefficients of tadpoles}
	
		All reduction coefficients vanish.
		
     \textbf{Reduction coefficient of bubbles}
	\bea
	C^{(0,1)}(1\vert 2)=\frac{s_{01} s_{12}-s_{02} s_{11}}{G(1,2)}.
	\eea
	Choosing $j_0=1,j_1=2$ in \eref{General-Coeff}, we have
	\bea
	C^{(1,2)}(1\vert 2)
	&=&\left.C^{(0,2)}(1\vert 2)\right\vert_{M_0\leftrightarrow M_1,K_1\to -K_1,K_2\to K_2-K_1}\nn
	&=&\left.\frac{s_{02} s_{12}-s_{01} s_{22}}{G(1,2)}\right\vert_{M_0\leftrightarrow M_1,K_1\to -K_1,K_2\to K_2-K_1}\nn
	&=&\frac{s_{02} \left(s_{11}-s_{12}\right)}{G(1,2)}+(1\leftrightarrow 2).
	\eea
	\textbf{Reduction coefficient of triangle}
	\bea
	C^{(0,1,2)}(1\vert 2)=\frac{ s_{01} \left(f _1 s_{22}-f _2 s_{12}\right)}{G(1,2)}+(1\leftrightarrow 2).
	\eea
	
	\item $m=2$
	
	\textbf{Reduction coefficients of tadpoles}
	\bea
	C^{(0)}(2\vert 2)=\frac{s_{11}s_{22}s_{01}s_{02}-s_{12}s_{22}s_{01}^2}{ s_{11} s_{22}G(1,2)}+(1\leftrightarrow 2).
	\eea
	Choose $j_0=1$ in \eref{General-Coeff}, we have
	\bea
	C^{(1)}(2\vert 2)&=&
	\left.C^{(0)}(2\vert 2)\right\vert_{M_0\leftrightarrow M_1,K_1\to -K_1,K_2\to K_2-K_1}\nn
	&=&\left.-\frac{ s_{12} s_{22} s_{01}^2-2 s_{02} s_{11} s_{22} s_{01}+s_{02}^2 s_{11} s_{12}}{ s_{11} s_{22} \left(s_{11} s_{22}-s_{12}^2\right)}\right\vert_{M_0\leftrightarrow M_1,K_1\to -K_1,K_2\to K_2-K_1}\nn
	&=&\frac{\left(-2 s_{12}^2+s_{22} s_{12}+s_{11} s_{22}\right) s_{01}^2+2 s_{02} s_{11} \left(s_{12}-s_{22}\right) s_{01}+s_{02}^2 s_{11} \left(s_{12}-s_{11}\right)}{s_{11} \left(s_{11}-2 s_{12}+s_{22}\right) G(1,2)}.\nn	\eea
	\textbf{Reduction coefficients of bubbles}
    \bea
    C^{(0,1)}(2\vert 2)=\frac{s_{01}^2 c_{2,0}^{(0,1)}(2)}{M_0^2}+\frac{s_{02} s_{01} c_{1,1}^{(0,1)}(2)}{M_0^2}+\frac{s_{02}^2 c_{0,2}^{(0,1)}(2)}{M_0^2}+s_{00} c_{0,0}^{(0,1)}(2),
    \eea
	where
	\bea
	c^{(0,1)}_{0,0}(2)=\frac{f_2 s_{11}-f_1 s_{12}}{(D-2)G(1,2)},
	\eea
	\bea
	c^{(0,1)}_{1,1}(2)=\frac{2 (D-1) f_2 M_0^2 s_{11} s_{12}}{(D-2) \left(s_{12}^2-s_{11} s_{22}\right){}^2}-\frac{2 f_1 M_0^2 \left((D-2) s_{11} s_{22}+s_{12}^2\right)}{(D-2) \left(s_{12}^2-s_{11} s_{22}\right){}^2}
	\eea
	\bea
	c^{(0,1)}_{2,0}(2)=
	-\frac{M_0^2 \left(f_2 s_{11} \left((D-2) s_{12}^2+s_{11} s_{22}\right)+f_1 s_{12} \left((D-2) s_{12}^2+(3-2 d) s_{11} s_{22}\right)\right)}{(D-2) s_{11} G(1,2)^2},\nn
	\eea
	\bea
	c^{(0,1)}_{0,2}(2)=-\frac{(D-1) M_0^2 s_{11} \left(f_2 s_{11}-f_1 s_{12}\right)}{(D-2) ^2G(1,2)^2},
	\eea
	 \bea
	C^{(1,2)}(2\vert 2)&=&\left.2 (2R\cdot K_2)\left [C^{(0,1)}(1,2)\right\vert_{\sigma}\right]+\left.C^{(0,1)}(2,2)\right\vert_{\sigma}\nn
	&=&4 s_{02}\left [\frac{ s_{01} \left(f _1 s_{22}-f _2 s_{12}\right)}{G(1,2)}+(1\leftrightarrow 2)\right ]+\left.C^{(0,1)}(2,2)\right\vert_{\sigma},\nn
	\eea
	where
	\bea
	\sigma=M_0\leftrightarrow M_2,K_2\to -K_2,K_1\to K_1-K_2.
	\eea
	\textbf{Reduction coefficients of triangle}
	\bea
	C^{(0,1,2)}(2\vert 2)=c^{(0,1,2)}_{0,0}M_0^2s_{00}+c^{(0,1,2)}_{2,0}s_{01}^2+c^{(0,1,2)}_{0,2}s_{02}^2+c^{(0,1,2)}_{1,1}s_{01}s_{02}.
	\eea
	where
	\bea
	c_{0,0}^{(0,1,2)}(2)=\frac{f_1^2 s_{22}-2 f_2 f_1 s_{12}+f_2^2 s_{11}+4 M_0^2 \left(s_{12}^2-s_{11} s_{22}\right)}{(D-2) M_0^2 \left(s_{12}^2-s_{11} s_{22}\right)},
	\eea
	\bea
	c_{2,0}^{(0,1,2)}(2)&=&\frac{f_2^2 \left((D-2) s_{12}^2+s_{11} s_{22}\right)}{(D-2) \left(s_{12}^2-s_{11} s_{22}\right)^2}-\frac{2 (D-1) f_1 f_2 s_{12} s_{22}}{(D-2) \left(s_{12}^2-s_{11} s_{22}\right)^2}\nn
	&&+\frac{(D-1) f_1^2 s_{22}^2}{(D-2) \left(s_{12}^2-s_{11} s_{22}\right)^2}+\frac{4 M_0^2 s_{22}}{(D-2) \left(s_{12}^2-s_{11} s_{22}\right)},
	\eea
	\bea
	c_{0,2}^{(0,1,2)}(2)&=&\frac{f_1^2 \left((D-2) s_{12}^2+s_{11} s_{22}\right)}{(D-2) \left(s_{12}^2-s_{11} s_{22}\right)^2}-\frac{2 (D-1) f_2 f_1 s_{11} s_{12}}{(D-2) \left(s_{12}^2-s_{11} s_{22}\right)^2}\nn
	&&+\frac{(D-1) f_2^2 s_{11}^2}{(D-2) \left(s_{12}^2-s_{11} s_{22}\right)^2}+\frac{4 M_0^2 s_{11}}{(D-2) \left(s_{12}^2-s_{11} s_{22}\right)},
	\eea
	\bea
	c_{1,1}^{(0,1,2)}(2)&=&-\frac{2 (D-1) f_1^2 s_{12} s_{22}}{(D-2) \left(s_{12}^2-s_{11} s_{22}\right)^2}+\frac{2 f_2 f_1 \left(D s_{12}^2+(D-2) s_{11} s_{22}\right)}{(D-2) \left(s_{12}^2-s_{11} s_{22}\right)^2}\nn
	&&-\frac{2 (D-1) f_2^2 s_{11} s_{12}}{(D-2) \left(s_{12}^2-s_{11} s_{22}\right)^2}-\frac{8 M_0^2 s_{12}}{(D-2) \left(s_{12}^2-s_{11} s_{22}\right)}.
	\eea

\end{itemize}
\subsection{All reduction coefficients of tensor box with rank $m=1,2$}
The MIs of a tensor box $I_4^{(m)}$ are
\bea
Tadpoles:&&I_1[0],I_1[1],I_1[2],I_1[3],\nn
Bubbles:&&I_2[0,1],I_2[0,2],I_2[0,3],I_2[1,2],I_2[1,3],I_2[2,3],\nn
Triangles:&&I_3[0,1,2],I_3[0,1,3],I_3[0,2,3],I_3[1,2,3],\nn
Box:&&I_4[0,1,2,3].
\eea
Then the reduction of the tensor box is
\bea
I_{4}^{(m)}&=&\sum_{i=0}^3C^{(i)}(m\vert 3)I_1[i]+\sum_{0\le i_1<i_2\le 3}C^{(i_1,i_2)}(m\vert 3)I_2[i_1,i_2]\nn
&&+\sum_{0\le i_1<i_2<i_3\le 3}C^{(i_1,i_2,i_3)}(m\vert 3)I_3[i_1,i_2,i_3]+C^{(0,1,2,3)}I_4[0,1,2,3].
\eea
\begin{itemize}
	\item $m=1$
	
	\textbf{Reduction coefficients of tadpoles, bubbles}
	
	All reduction coefficients vanish.
	
	\textbf{Reduction coefficients of triangles}
	\bea
	C^{(0,1,2)}(1\vert 3)=-\frac{G(2,3;1,2) s_{01}-G(1,3;1,2) s_{02}+G(1,2;1,2) s_{03}}{ G(1,2,3)}.
	\eea
	In \eref{General-Coeff}, choosing $j_0=3,j_1=1,j_2=2$, we have
	\bea
	C^{(1,2,3)}(1\vert 3)
	&=&C^{(3,1,2)}(1\vert 3)=(2R\cdot K_3)\left.C^{(0,1,2)}(0\vert 3)\right\vert_{K_1\to K_1-K_3,K_2\to K_2-K_3,K_3\to -K_3,M_0\leftrightarrow M_3}\nn
	&&+\left.C^{(0,1,2)}(1\vert 3)\right\vert_{K_1\to K_1-K_3,K_2\to K_2-K_3,K_3\to -K_3,M_0\leftrightarrow M_3}\nn
	&=&\frac{G(K_2-K_3,K_3;K_1-K_3,K_2-K_3) (s_{01}-s_{03})}{ G(K_1-K_3,K_2-K_3,K_3;K_1-K_3,K_2-K_3,K_3)}\nn
	&&+\frac{-G(K_1-K_3,K_3;K_1-K_3,K_2-K_3) (s_{02}-s_{03})}{ G(K_1-K_3,K_2-K_3,K_3;K_1-K_3,K_2-K_3,K_3)}\nn
	&&+\frac{G(K_1-K_3,K_2-K_3;K_1-K_3,K_2-K_3) s_{03}}{ G(K_1-K_3,K_2-K_3,K_3;K_1-K_3,K_2-K_3,K_3)}.
	\eea
	\textbf{Reduction coefficients of box}
	\bea
	C^{(0,1,2,3)}(1\vert 3)&=&\frac{f_3 \left(s_{01} G(2,3;1,2)-s_{02} G(1,3;1,2)+s_{03} G(1,2;1,2)\right)}{G(1,2,3)}\nn
	&&-\frac{f_2 \left(s_{01} G(2,3;1,3)-s_{02} G(1,3;1,3)+s_{03} G(1,3;1,2)\right)}{G(1,2,3)}\nn
	&&+\frac{f_1 \left(s_{01} G(2,3;2,3)-s_{02} G(2,3;1,3)+s_{03} G(2,3;1,2)\right)}{G(1,2,3)}.
	\eea
	\item $m=2$
	
	\textbf{Reduction coefficients of tadpoles}
	
	All reduction coefficients vanish.
	
	\textbf{Reduction coefficients of bubbles}
	\bea
	C^{(0,1)}(2\vert 3)&=&(M_0^2)^{-2}\Big[c_{0,0,0}^{(0,1)}(2)M_0^2s_{00}+c_{2,0,0}^{(0,1)}(2)s_{01}^2+c_{0,2,0}^{(0,1)}(2)s_{02}^2+c_{0,0,2}^{(0,1)}(2)s_{03}^2\nn
	&&+c_{1,1,0}^{(0,1)}(2)s_{01}s_{02}+c_{0,1,1}^{(0,1)}(2)s_{02}s_{03}+c_{1,0,1}^{(0,1)}(2)s_{01}s_{03}\Big].
	\eea
	where
	\bea
	c_{0,0,0}^{(0,1)}(2)=0,
	\eea
	\bea
	c_{2,0,0}^{(0,1)}(2)=\frac{M_0^4 s_{13} G(2,3;1,3)}{G(1,3;1,3) G(1,2,3)}-\frac{M_0^4 s_{12} G(2,3;1,2)}{G(1,2;1,2) G(1,2,3)},
	\eea
	\bea
	c_{0,2,0}^{(0,1)}(2)&=&-\frac{M_0^4 s_{11} G(1,3;1,2)}{G(1,2;1,2) G(1,2,3)},\nn
	c_{0,0,2}^{(0,1)}(2)&=&\left.c_{0,2,0}^{(0,1)}(2)\right\vert_{2\leftrightarrow 3},
	\eea
	\bea
	c_{0,1,1}^{(0,1)}(2)=\frac{2 M_0^4 s_{11}}{G(1,2,3)},
	\eea
	\bea
	c_{1,1,0}^{(0,1)}(2)&=&\frac{2 M_0^4 s_{11} G(2,3;1,2)}{G(1,2;1,2) G(1,2,3)},\nn
	c_{1,0,1}^{(0,1)}(2)&=&\left.c_{1,1,0}^{(0,1)}(2)\right\vert_{2\leftrightarrow 3}.
	\eea
   Choosing $j_0=2,j_1=1$ in \eref{General-Coeff} we have
	\bea
	C^{(1,2)}(2\vert 3)&=&\left  [\left. 2(-2R\cdot K_2)C^{(0,1)}(1\vert 3)+C^{(0,1)}(2\vert 3)\right  ]\right\vert_{K_1\to K_1-K_2,K_2 \to -K_2,K_3\to K_3-K_2 M_0\leftrightarrow
		M_2}\nn
	&=&\left. C^{(0,1)}(2\vert 3)\right\vert_{K_1\to K_1-K_2,K_2 \to -K_2,K_3\to K_3-K_2 M_0\leftrightarrow
		M_2}.
	\eea
	\textbf{Reduction coefficients of triangles}
	\bea
	C^{(0,1,2)}(2\vert 3)&=&M_0^{-2}\Big[c_{0,0,0}^{(0,1,2)}(2)M_0^2s_{00}+c_{2,0,0}^{(0,1,2)}(2)s_{01}^2+c_{0,2,0}^{(0,1,2)}(2)s_{02}^2+c_{0,0,2}^{(0,1,2)}(2)s_{03}^2\nn
	&&+c_{1,1,0}^{(0,1,2)}(2)s_{01}s_{02}+c_{0,1,1}^{(0,1,2)}(2)s_{02}s_{03}+c_{1,0,1}^{(0,1,2)}(2)s_{01}s_{03}\Big],
	\eea
	where
	\bea
	c_{0,0,0}^{(0,1,2)}(2)&=&\frac{f_1 G(2,3;1,2)}{(D-3) G(1,2,3)}-\frac{f_2 G(1,3;1,2)}{(D-3) G(1,2,3)}+\frac{f_3 G(1,2;1,2)}{(D-3) G(1,2,3)},
	\eea
	\bea
	c_{2,0,0}^{(0,1,2)}(2)&=&\frac{f_2 M_0^2 G(1,3;1,2) G(2,3;2,3)}{(D-3) G(1,2,3)^2}-\frac{f_3 M_0^2 G(1,2;1,2) G(2,3;2,3)}{(D-3) G(1,2,3)^2}\nn
	&&+G(2,3;1,2) \left(\frac{f_2 M_0^2 G(2,3;1,3)}{G(1,2,3)^2}-\frac{(D-2) f_1 M_0^2 G(2,3;2,3)}{(D-3) G(1,2,3)^2}\right)\nn
	&&+\frac{M_0^2 \left(f_2 s_{12}-f_1 s_{22}\right) G(2,3;1,2)}{G(1,2;1,2) G(1,2,3)}-\frac{f_3 M_0^2 G(2,3;1,2)^2}{G(1,2,3)^2},
	\eea
	\bea
	c_{0,2,0}^{(0,1,2)}(2)&=&\left.c_{2,0,0}^{(0,1,2)}(2)\right\vert_{1\leftrightarrow 2},
	\eea
	\bea
	c_{0,0,2}^{(0,1,2)}(2)&=&G(1,2;1,2) \left [\frac{(D-2) f_2 M_0^2 G(1,3;1,2)}{(D-3) G(1,2,3;1,2,3)^2}-\frac{(D-2) f_1 M_0^2 G(2,3;1,2)}{(D-3) G(1,2,3;1,2,3)^2}\right]\nn
	&&-\frac{(D-2) f_3 M_0^2 G(1,2;1,2)^2}{(D-3) G(1,2,3;1,2,3)^2},
	\eea
	\bea
	c_{1,1,0}^{(0,1,2)}(2)&=&-\frac{2 (D-2) f_2 M_0^2 G(1,3;1,2) G(2,3;1,3)}{(D-3) G(1,2,3)^2}+\frac{2 f_3 M_0^2 G(1,2;1,2) G(2,3;1,3)}{(D-3) G(1,2,3)^2}\nn
	&&+G(2,3;1,2) \left(\frac{2 f_1 M_0^2 G(2,3;1,3)}{(D-3) G(1,2,3)^2}+\frac{2 f_3 M_0^2 G(1,3;1,2)}{G(1,2,3)^2}\right)\nn
	&&+\frac{2 M_0^2 \left(f_1 s_{12}-f_2 s_{11}\right) G(2,3;1,2)}{G(1,2;1,2) G(1,2,3)}+\frac{2 f_1 M_0^2 G(1,3;1,2) G(2,3;2,3)}{G(1,2,3)^2},\nn
	\eea
	\bea
	c_{0,1,1}^{(0,1,2)}(2)&=&\frac{2 f_1 M_0^2 G(1,3;1,2) G(2,3;1,2)}{(D-3) G(1,2,3)^2}-\frac{2 f_2 M_0^2 G(1,3;1,2)^2}{(D-3) G(1,2,3)^2}\nn
	&&+G(1,2;1,2) \left(\frac{2 (D-2) f_3 M_0^2 G(1,3;1,2)}{(D-3) G(1,2,3)^2}+\frac{2 f_1 M_0^2 G(2,3;1,3)}{G(1,2,3)^2}\right)\nn
	&&-\frac{2 f_2 M_0^2 G(1,2;1,2)G(1,3;1,3)}{G(1,2,3)^2},
	\eea
	\bea
	c_{1,0,1}^{(0,1,2)}(2)=\left.c_{0,1,1}^{(0,1,2)}(2)\right\vert_{1\leftrightarrow 2}.
	\eea

	Choosing $j_0=3,j_1=1,j_2=2$ in  \eref{General-Coeff}, we have
	\bea
	C^{(1,2,3)}(1\vert 3)&=&\left.\left [2(-2R\cdot K_3)C^{(0,1,2)}(1\vert 3)+C^{(0,1,2)}(2\vert 3)\right]\right\vert_{K_1\to K_1-K_3,K_2\to K_2-K_3,K_3\to -K_3,M_0\leftrightarrow M_3}\nn
	&=&4s_{03}\Bigg \{\frac{G(K_2-K_3,K_3;K_1-K_3,K_2-K_3) (s_{01}-s_{03})}{ G(K_1-K_3,K_2-K_3,K_3;K_1-K_3,K_2-K_3,K_3)}\nn
	&&+\frac{-G(K_1-K_3,K_3;K_1-K_3,K_2-K_3) (s_{02}-s_{03})}{ G(K_1-K_3,K_2-K_3,K_3;K_1-K_3,K_2-K_3,K_3)}\nn
	&&+\frac{G(K_1-K_3,K_2-K_3;K_1-K_3,K_2-K_3) s_{03}}{ G(K_1-K_3,K_2-K_3,K_3;K_1-K_3,K_2-K_3,K_3)}\Bigg\}\nn
	&&+\left.C^{(0,1,2)}(2,3)\right\vert_{K_1\to K_1-K_3,K_2\to K_2-K_3,K_3\to -K_3,M_0\leftrightarrow M_3}.
	\eea
	\textbf{Reduction coefficient of box}
	\bea
	C^{(0,1,2,3)}(2\vert 3)&=&c_{0,0,0}^{(0,1,2,3)}(2)M_0^2s_{00}+c_{2,0,0}^{(0,1,2,3)}(2)s_{01}^2+c_{0,2,0}^{(0,1,2,3)}(2)s_{02}^2+c_{0,0,2}^{(0,1,2,3)}(2)s_{03}^2\nn
	&&+c_{1,1,0}^{(0,1,2,3)}(2)s_{01}s_{02}+c_{0,1,1}^{(0,1,2,3)}(2)s_{02}s_{03}+c_{1,0,1}^{(0,1,2,3)}(2)s_{01}s_{03},
	\eea
	where
	\bea
	c_{0,0,0}^{(0,1,2,3)}(2)&=&-\frac{f_1^2 G(2,3;2,3)}{(D-3) M_0^2 G(1,2,3)}+\frac{2 f_2 f_1 G(2,3;1,3)}{(D-3) M_0^2 G(1,2,3)}\nn
	&&-\frac{2 f_3 f_1 G(2,3;1,2)}{(D-3) M_0^2 G(1,2,3)}-\frac{f_3^2 G(1,2;1,2)}{(D-3) M_0^2 G(1,2,3)}\nn
	&&+\frac{f_2 \left(2 f_3 G(1,3;1,2)-f_2 G(1,3;1,3)\right)}{(D-3) M_0^2 G(1,2,3)}+\frac{4}{D-3},
	\eea
	\bea
	c_{2,0,0}^{(0,1,2,3)}(2)&=&
	\frac{f_3^2 G(1,2;1,2)G(2,3;2,3)}{(D-3) G(1,2,3)^2}+\frac{2 (D-2) f_1 f_3 G(2,3;1,2)G(2,3;2,3)}{(D-3) G(1,2,3)^2}\nn
	&&-\frac{2 (D-2) f_1 f_2 G(2,3;1,3)G(2,3;2,3)}{(D-3) G(1,2,3)^2}+\frac{f_2^2 G(2,3;1,3)^2}{G(1,2,3)^2}\nn
	&&+\frac{f_2 \left(f_2 G(1,3;1,3)G(2,3;2,3)-2 f_3 G(1,3;1,2)G(2,3;2,3)\right)}{(D-3) G(1,2,3)^2}\nn
	&&-\frac{4 M_0^2G(2,3;2,3)}{(D-3) G(1,2,3)}+\frac{(D-2) f_1^2 G(2,3;2,3)^2}{(D-3) G(1,2,3)^2}\nn
	&&+\frac{f_3^2 G(2,3;1,2)^2}{G(1,2,3)^2}-\frac{2 f_2 f_3 G(2,3;1,2) G(2,3;1,3)}{G(1,2,3)^2},
	\eea
	\bea
	&&c_{1,1,0}^{(0,1,2,3)}(2)=\frac{8 M_0^2G(2,3;1,3)}{(D-3) G(1,2,3)}\nn
	&&-\frac{2 f_2 \left((D-2) f_2 G(1,3;1,3)G(2,3;1,3)-(D-1) f_3 G(1,3;1,2)G(2,3;1,3)\right)}{(D-3) G(1,2,3)^2}\nn
	&&+\frac{2 (D-1) f_1 f_2 G(2,3;1,3)^2}{(D-3) G(1,2,3)^2}-\frac{2 f_3^2 G(1,2;1,2) G(2,3;1,3)}{(D-3) G(1,2,3)^2}\nn
	&&+ \frac{2 f_1G(2,3;2,3) \left(f_2 G(1,3;1,3)-f_3 G(1,3;1,2)\right)}{G(1,2,3)^2}\nn
	&&-\frac{2 (D-2) f_1^2 G(2,3;1,3)G(2,3;2,3)}{(D-3) G(1,2,3)^2}-\frac{2 (D-1) f_1 f_3 G(2,3;1,3)G(2,3;1,2)}{(D-3) G(1,2,3)^2}\nn
	&&+ \frac{2 f_3 G(2,3;1,2)\left(f_2 G(1,3;1,3)-f_3 G(1,3;1,2)\right)}{G(1,2,3)^2}.
	\eea
	Other expansion coefficients can be got by using the permutation symmetry:
	\bea
	c_{0,2,0}^{(0,1,2,3)}(2)&=&\left.c_{2,0,0}^{(0,1,2,3)}(2)\right\vert_{1\leftrightarrow 2},\nn
	c_{0,0,2}^{(0,1,2,3)}(2)&=&\left.c_{2,0,0}^{(0,1,2,3)}(2)\right\vert_{1\leftrightarrow 3},\nn
	c_{1,0,1}^{(0,1,2,3)}(2)&=&\left.c_{1,1,0}^{(0,1,2,3)}(2)\right\vert_{2\leftrightarrow 3},\nn
	c_{0,1,1}^{(0,1,2,3)}(2)&=&\left.c_{1,1,0}^{(0,1,2,3)}(2)\right\vert_{1\leftrightarrow 3}.
	\eea
\end{itemize}
\subsection{All reduction coefficients of tensor pentagon with rank $m=1,2$}
Consider the reduction of a tensor pentagon $I_5^{(m)}$, the MIs are as below
\bea
Tadpoles:&&I_1[0],I_1[1],I_1[2],I_1[3],I_1[4],\nn
Bubbles:&&I_2[0,1],I_2[0,2],I_2[0,3],I_2[0,4],I_2[1,2],I_2[1,3],I_2[1,4],I_2[2,3],I_2[2,4],I_2[3,4],\nn
Triangles:&&I_3[0,1,2],I_3[0,1,3],I_3[0,1,4],I_3[0,2,3],I_3[0,2,4],I_3[0,3,4],I_3[1,2,3],\nn
&&I_3[1,2,4],I_3[1,3,4],I_3[2,3,4],\nn
Box:&&I_4[0,1,2,3],I_4[0,2,3,4],I_4[0,1,2,4],I_4[1,2,3,4],\nn
Pentagon:&& I_5[0,1,2,3,4].
\eea
Then the reduction of the tensor pentagon is given by
\bea
I_{5}^{(m)}&=&\sum_{i=0}^4C_0^{(i)}(m\vert 4)I_1[i]+\sum_{0\le i_1<i_2\le 4}C^{(i_1,i_2)}(m\vert 4)I_2[i_1,i_2]+\sum_{0\le i_1<i_2<i_3\le 4}C^{(i_1,i_2,i_3)}(m\vert 4)I_3[i_1,i_2,i_3]\nn
&&+\sum_{0\le i_1<i_2<i_3<i_4\le 4}C^{(i_1,i_2,i_3,i_4)}(m\vert 4)I_4[i_1,i_2,i_3,i_4]+C^{(0,1,2,3,4)}(m\vert 4)I_5[0,1,2,3,4].
\eea
\begin{itemize}
	\item $m=1$
	
	\textbf{Reduction coefficients of tadpoles, bubbles, triangles}
	
	All reduction coefficients vanish.

	\textbf{Reduction coefficients of boxes}
	\bea
	C^{(0,1,2,3)}(1\vert 4)&=&\frac{s_{01} G(2,3,4;1,2,3)-s_{02} G(1,3,4;1,2,3)+s_{03} G(1,2,4;1,2,3)}{G(1,2,3,4;1,2,3,4)}\nn
	&&-\frac{s_{04} G(1,2,3)}{G(1,2,3,4;1,2,3,4)}.
	\eea
    Choosing $j_0=4,j_1=1,j_2=2,j_3=3$ in \eref{General-Coeff}, we have
	\bea
	C^{(1,2,3,4)}(1\vert 4)
	&=&\left.C^{(0,1,2,3)}(1\vert 4)\right\vert_{K_1\to K_1-K_4,K_2\to K_2-K_4,K_3\to K_3-K_4,K_4\to -K_4,M_0\leftrightarrow M_4}\nn
	&=&\frac{(-s_{01}+s_{04}) G(K_2-K_4,K_3-K_4,K_4;K_1-K_4,K_2-K_4,K_3-K_4)}{G(K_1-K_4,K_2-K_4,K_3-K_4,K_4;K_1-K_4,K_2-K_4,K_3-K_4,K_4)}\nn
	&&+\frac{(s_{02}-s_{04}) G(K_1-K_4,K_3-K_4,K_4;K_1-K_4,K_2-K_4,K_3-K_4)}{G(K_1-K_4,K_2-K_4,K_3-K_4,K_4;K_1-K_4,K_2-K_4,K_3-K_4,K_4)}\nn
	&&-\frac{(s_{03}-s_{04}) G(K_1-K_4,K_2-K_4,K_4;K_1-K_4,K_2-K_4,K_3-K_4)}{G(K_1-K_4,K_2-K_4,K_3-K_4,K_4;K_1-K_4,K_2-K_4,K_3-K_4,K_4)}\nn
	&&+\frac{s_{04} G(K_1-K_4,K_2-K_4,K_3-K_4;K_1-K_4,K_2-K_4,K_3-K_4)}{G(K_1-K_4,K_2-K_4,K_3-K_4,K_4;K_1-K_4,K_2-K_4,K_3-K_4,K_4)}.\nn
	\eea
	
	\textbf{Reduction coefficients of pentagon}
	\bea
	C^{(0,1,2,3,4)}(1\vert 4)=c_{1,0,0,0}^{(0,1,2,3,4)}(1)s_{01}+c_{0,1,0,0}^{(0,1,2,3,4)}(1)s_{02}+c_{0,0,1,0}^{(0,1,2,3,4)}(1)s_{03}+c_{0,0,0,1}^{(0,1,2,3,4)}(1)s_{04},\nn
	\eea
	where
	\bea
	c_{1,0,0,0}^{(0,1,2,3,4)}(1)&=&\frac{f_1 G(2,3,4)-f_2 G(2,3,4;1,3,4)+f_3 G(2,3,4;1,2,4)}{G(1,2,3,4;1,2,3,4)}\nn
	&&-\frac{f_4 G(2,3,4;1,2,3)}{G(1,2,3,4;1,2,3,4)},
	\eea
	\bea
	c_{0,1,0,0}^{(0,1,2,3,4)}(1)&=&\left.c_{1,0,0,0}^{(0,1,2,3,4)}(1)\right\vert_{1\leftrightarrow 2},\nn
	c_{0,0,1,0}^{(0,1,2,3,4)}(1)&=&\left.c_{1,0,0,0}^{(0,1,2,3,4)}(1)\right\vert_{1\leftrightarrow 3},\nn
	c_{0,0,0,1}^{(0,1,2,3,4)}(1)&=&\left.c_{1,0,0,0}^{(0,1,2,3,4)}(1)\right\vert_{1\leftrightarrow 4}.
	\eea
	\item $m=2$
	
	\textbf{Reduction coefficients of tadpoles, bubbles}
	
	All reduction coefficients vanish.

	\textbf{Reduction coefficients of triangles}
	\bea
	C^{(0,1,2)}(2\vert 4)&=&-\frac{s_{04}^2 G(1,2;1,2) G(1,2,4;1,2,3)}{G(1,2,4;1,2,4) G(1,2,3,4;1,2,3,4)}-\frac{s_{03}^2 G(1,2;1,2) G(1,2,4;1,2,3)}{G(1,2,3)G(1,2,3,4;1,2,3,4)}\nn
	%
	&&-\frac{s_{01} G(2,3,4;1,2,3) \left(s_{01} G(2,3;1,2)-2 s_{02} G(1,3;1,2)+2 s_{03} G(1,2;1,2)\right)}{G(1,2,3) G(1,2,3,4;1,2,3,4)}\nn
	&&+\frac{s_{02} G(1,3,4;1,2,3) \left(2 s_{03} G(1,2;1,2)-s_{02} G(1,3;1,2)\right)}{G(1,2,3)G(1,2,3,4;1,2,3,4)}\nn
	&&+\frac{s_{01}^2 G(2,4;1,2) G(2,3,4;1,2,4)}{G(1,2,4;1,2,4) G(1,2,3,4;1,2,3,4)}\nn
	&&+\frac{s_{02} G(1,4;1,2) \left(s_{02} G(1,3,4;1,2,4)-2 s_{01} G(2,3,4;1,2,4)\right)}{G(1,2,4;1,2,4) G(1,2,3,4;1,2,3,4)}\nn
	&&+\frac{2 s_{04} G(1,2;1,2) \left(s_{01} G(2,3,4;1,2,4)-s_{02} G(1,3,4;1,2,4)+s_{03} G(1,2,4;1,2,4)\right)}{G(1,2,4;1,2,4) G(1,2,3,4;1,2,3,4)}.\nn
	\eea
	Choosing $j_0=3,j_1=1,j_2=2 $ in  \eref{General-Coeff}, we have
	\bea
	C^{(1,2,3)}(2\vert 4)
	&=&\left.\left (C^{(0,1,2)}(2\vert 4)\right)\right\vert_{K_i\to K_i-K_3,i\not=3;K_3\to -K_3,M_0\leftrightarrow M_3}.
	\eea

	\textbf{Reduction coefficients of boxes}
	\bea
	C^{(0,1,2,3)}(2\vert 4)&=&{1\over M_0^2}\Bigg [c_{0,0,0,0}^{(0,1,2,3)}(2)M_0^2s_{00}+c_{2,0,0,0}^{(0,1,2,3)}(2)s_{01}^2+c_{0,2,0,0}^{(0,1,2,3)}(2)s_{02}^2+c_{0,0,2,0}^{(0,1,2,3)}(2)s_{03}^2\nn
	&&+c_{0,0,0,2}^{(0,1,2,3)}(2)s_{04}^2+c_{1,1,0,0}^{(0,1,2,3)}(2)s_{01}s_{02}+c_{0,1,1,0}^{(0,1,2,3)}(2)s_{02}s_{03}+c_{1,0,1,0}^{(0,1,2,3)}(2)s_{01}s_{03}\nn
	&&+c_{1,0,0,1}^{(0,1,2,3)}(2)s_{01}s_{04}+c_{0,1,0,1}^{(0,1,2,3)}(2)s_{02}s_{04}+c_{0,0,1,1}^{(0,1,2,3)}(2)s_{03}s_{04}\Bigg].
	\eea
	where
	\bea
	c_{0,0,0,0}^{(0,1,2,3)}(2)&=&\frac{f_1 G(2,3,4;1,2,3)}{(4-D) G(1,2,3,4)}+\frac{f_2 G(1,3,4;1,2,3)}{(D-4) G(1,2,3,4)}\nn
	&&+\frac{f_3 G(1,2,4;1,2,3)}{(4-D) G(1,2,3,4)}+\frac{f_4 G(1,2,3)}{(D-4) G(1,2,3,4)}
	\eea
	\bea
	c_{2,0,0,0}^{(0,1,2,3)}(2)&=&\frac{(D-3) f_1 M_0^2 G(2,3,4;1,2,3)G(2,3,4)}{(D-4) G(1,2,3,4;1,2,3,4)^2}\nn
	&&-\frac{M_0^2 G(2,3,4)\left(f_2 G(1,3,4;1,2,3)-f_3 G(1,2,4;1,2,3)\right)}{(D-4) G(1,2,3,4;1,2,3,4)^2}\nn
	&&-\frac{f_4 M_0^2 G(1,2,3)G(2,3,4)}{(D-4) G(1,2,3,4;1,2,3,4)^2}
	-\frac{f_4 M_0^2 G(2,3,4;1,2,3)^2}{G(1,2,3,4;1,2,3,4)^2}\nn
	&&+G(2,3,4;1,2,3)\frac{M_0^2 \left(f_1 G(2,3;2,3)-f_2 G(2,3;1,3)+f_3 G(2,3;1,2)\right)}{G(1,2,3) G(1,2,3,4;1,2,3,4)}\nn
	&&-G(2,3,4;1,2,3)\frac{M_0^2 \left(f_2 G(2,3,4;1,3,4)-f_3 G(2,3,4;1,2,4)\right)}{G(1,2,3,4;1,2,3,4)^2},
	\eea
	\bea
	c_{0,0,0,2}^{(0,1,2,3)}(2)&=&\frac{(D-3) M_0^2 G(1,2,3) }{(D-4) G(1,2,3,4;1,2,3,4)^2}\Big [f_1 (G(2,3,4;1,2,3))\nn
	&&-f_2 G(1,3,4;1,2,3)+f_3 G(1,2,4;1,2,3)-f_4 G(1,2,3)\Big],
	\eea
	\bea
	c_{1,1,0,0}^{(0,1,2,3)}(2)&=&\frac{2 (D-3) f_2 M_0^2 G(1,3,4;1,2,3) G(2,3,4;1,3,4)}{(D-4) G(1,2,3,4;1,2,3,4)^2}\nn
	&&+\frac{2 M_0^2  G(2,3,4;1,3,4)}{(D-4) G(1,2,3,4;1,2,3,4)^2}\Big [f_4  G(1,2,3)- f_3 G(1,2,4;1,2,3)\Big]\nn
	&&+{2M_0^2G(2,3,4;1,2,3)\over G(1,2,3,4;1,2,3,4)^2 } \Big[f_4  G(1,3,4;1,2,3)-\frac{f_1 G(2,3,4;1,3,4)}{(D-4)}\Big]\nn
	&&-\frac{2 M_0^2 G(1,3,4;1,2,3) }{G(1,2,3,4;1,2,3,4)^2}\Big[f_1 G(2,3,4)- f_3  G(2,3,4;1,2,4)\Big]\nn
	&&-\frac{2 M_0^2 G(2,3,4;1,2,3) \left(f_1 G(2,3;1,3)-f_2 G(1,3;1,3)+f_3 G(1,3;1,2)\right)}{G(1,2,3) G(1,2,3,4;1,2,3,4)},\nn
	\eea
	\bea
	c_{1,0,0,1}^{(0,1,2,3)}(2)&=&-\frac{2 f_1 M_0^2 G(2,3,4;1,2,3)^2}{(D-4) G(1,2,3,4;1,2,3,4)^2}\nn
	&&+{2M_0^2G(2,3,4;1,2,3) \over (D-4) G(1,2,3,4;1,2,3,4)^2}\Big [f_2 G(1,3,4;1,2,3)- f_3  G(1,2,4;1,2,3)\Big ]\nn
	&&+{2M_0^2G(1,2,3) \over  G(1,2,3,4;1,2,3,4)^2} \Bigg[\frac{(D-3) f_4G(2,3,4;1,2,3)}{(D-4) }\nn
	&&-f_1  G(2,3,4)+ f_2  G(2,3,4;1,3,4)- f_3  G(2,3,4;1,2,4)\Bigg].
	\eea
	Other expansion coefficients can be got by using the permutation symmetry:
	\bea
	c_{0,2,0,0}^{(0,1,2,3)}(2)&=&\left.c_{2,0,0,0}^{(0,1,2,3)}(2)\right\vert_{1\leftrightarrow 2},\nn
	c_{0,0,2,0}^{(0,1,2,3)}(2)&=&\left.c_{2,0,0,0}^{(0,1,2,3)}(2)\right\vert_{1\leftrightarrow 3},\nn
	c_{1,0,1,0}^{(0,1,2,3)}(2)&=&\left.c_{1,1,0,0}^{(0,1,2,3)}(2)\right\vert_{2\leftrightarrow 3},\nn
	c_{0,1,1,0}^{(0,1,2,3)}(2)&=&\left.c_{1,1,0,0}^{(0,1,2,3)}(2)\right\vert_{1\leftrightarrow 2},\nn
	c_{0,1,0,1}^{(0,1,2,3)}(2)&=&\left.c_{1,0,0,1}^{(0,1,2,3)}(2)\right\vert_{1\leftrightarrow 2},\nn
	c_{0,0,1,1}^{(0,1,2,3)}(2)&=&\left.c_{1,0,0,1}^{(0,1,2,3)}(2)\right\vert_{1\leftrightarrow 3}.
	\eea
   Choosing $j_0=4,j_1=1,j_2=2,j_3=3$ in \eref{General-Coeff}, we have
	\bea
	C^{(1,2,3,4)}(2\vert 4)
	&=&4 s_{04}\left.\left (C^{(0,1,2,3)}(1\vert 4)\right)\right\vert_{K_i\to K_i-K_4,i<4;K_4\to -K_4,M_0\leftrightarrow M_4}\nn
	&&+\left.\left (C^{(0,1,2,3)}(2\vert 4)\right)\right\vert_{K_i\to K_i-K_4,i<4;K_4\to -K_4,M_0\leftrightarrow M_4}.
	\eea

	\textbf{The reduction coefficient of pentagon}
	\bea
	C^{(0,1,2,3,4)}(2\vert 4)&=&\Bigg [c_{0,0,0,0}^{(0,1,2,3,4)}(2)M_0^2s_{00}+c_{2,0,0,0}^{(0,1,2,3,4)}(2)s_{01}^2+c_{0,2,0,0}^{(0,1,2,3,4)}(2)s_{02}^2+c_{0,0,2,0}^{(0,1,2,3,4)}(2)s_{03}^2\nn
	&&+c_{0,0,0,2}^{(0,1,2,3,4)}(2)s_{04}^2+c_{1,1,0,0}^{(0,1,2,3,4)}(2)s_{01}s_{02}+c_{0,1,1,0}^{(0,1,2,3,4)}(2)s_{02}s_{03}+c_{1,0,1,0}^{(0,1,2,3,4)}(2)s_{01}s_{03}\nn
	&&+c_{1,0,0,1}^{(0,1,2,3,4)}(2)s_{01}s_{04}+c_{0,1,0,1}^{(0,1,2,3,4)}(2)s_{02}s_{04}+c_{0,0,1,1}^{(0,1,2,3,4)}(2)s_{03}s_{04}\Bigg].
	\eea
	where
	\bea
	c_{0,0,0,0}^{(0,1,2,3,4)}(2)&=&-\frac{f_1^2 G(2,3,4)}{(D-4) M_0^2 G(1,2,3,4;1,2,3,4)}+\frac{2 f_2 f_1 G(2,3,4;1,3,4)}{(D-4) M_0^2 G(1,2,3,4;1,2,3,4)}\nn
	&&-\frac{2 f_3 f_1 G(2,3,4;1,2,4)}{(D-4) M_0^2 G(1,2,3,4;1,2,3,4)}+\frac{2 f_4 f_1 G(2,3,4;1,2,3)}{(D-4) M_0^2 G(1,2,3,4;1,2,3,4)}\nn
	&&-\frac{f_2^2 G(1,3,4)}{(D-4) M_0^2 G(1,2,3,4;1,2,3,4)}-\frac{f_3^2 G(1,2,4)}{(D-4) M_0^2 G(1,2,3,4;1,2,3,4)}\nn
	&&-\frac{f_4^2 G(1,2,3)}{(D-4) M_0^2 G(1,2,3,4;1,2,3,4)}+\frac{2 f_2 f_3 G(1,3,4;1,2,4)}{(D-4) M_0^2 G(1,2,3,4;1,2,3,4)}\nn
	&&-\frac{2 f_2 f_4 G(1,3,4;1,2,3)}{(D-4) M_0^2 G(1,2,3,4;1,2,3,4)}+\frac{2 f_3 f_4 G(1,2,4;1,2,3)}{(D-4) M_0^2 G(1,2,3,4;1,2,3,4)}\nn
	&&+\frac{4}{D-4},
	\eea
	\bea
	c^{(0,1,2,3,4)}_{2,0,0,0}(2)&=&\frac{(D-3) f_1^2 G(2,3,4)^2}{(D-4) G(1,2,3,4;1,2,3,4)^2}-\frac{2 (D-3) f_2 f_1 G(2,3,4;1,3,4) G(2,3,4)}{(D-4) G(1,2,3,4;1,2,3,4)^2}\nn
	&&+\frac{2 (D-3) f_3 f_1 G(2,3,4;1,2,4) G(2,3,4)}{(D-4) G(1,2,3,4;1,2,3,4)^2}-\frac{2 (D-3) f_4 f_1 G(2,3,4;1,2,3) G(2,3,4)}{(D-4) G(1,2,3,4;1,2,3,4)^2}\nn
	&&+f_2^2 \left(\frac{G(1,3,4) G(2,3,4)}{(D-4) G(1,2,3,4;1,2,3,4)^2}+\frac{G(2,3,4;1,3,4)^2}{G(1,2,3,4;1,2,3,4)^2}\right)\nn
	&&+\frac{f_3^2 \left((D-4) G(2,3,4;1,2,4)^2+G(1,2,4;1,2,4) G(2,3,4)\right)}{(D-4) G(1,2,3,4;1,2,3,4)^2}\nn
	&&+f_4^2 \left(\frac{G(1,2,3) G(2,3,4)}{(D-4) G(1,2,3,4;1,2,3,4)^2}+\frac{G(2,3,4;1,2,3)^2}{G(1,2,3,4;1,2,3,4)^2}\right)\nn
	&&+f_2 f_3 \left(-\frac{2 G(1,3,4;1,2,4) G(2,3,4)}{(D-4) G(1,2,3,4;1,2,3,4)^2}-\frac{2 G(2,3,4;1,2,4) G(2,3,4;1,3,4)}{G(1,2,3,4;1,2,3,4)^2}\right)\nn
	&&+f_2 f_4 \left(\frac{2 G(1,3,4;1,2,3) G(2,3,4)}{(D-4) G(1,2,3,4;1,2,3,4)^2}+\frac{2 G(2,3,4;1,2,3) G(2,3,4;1,3,4)}{G(1,2,3,4;1,2,3,4)^2}\right)\nn
	&&+f_3 f_4 \left(-\frac{2 G(1,2,4;1,2,3) G(2,3,4)}{(D-4) G(1,2,3,4;1,2,3,4)^2}-\frac{2 G(2,3,4;1,2,3) G(2,3,4;1,2,4)}{G(1,2,3,4;1,2,3,4)^2}\right)\nn
	&&-\frac{4 M_0^2 G(2,3,4)}{(D-4) G(1,2,3,4;1,2,3,4)},
	\eea
	\bea
	c^{(0,1,2,3,4)}_{1,1,0,0}(2)&=&-\frac{2 (D-3) G(2,3,4;1,3,4) G(2,3,4) f_1^2}{(D-4) G(1,2,3,4;1,2,3,4)^2}+\frac{8 G(2,3,4;1,3,4) M_0^2}{(D-4) G(1,2,3,4;1,2,3,4)}\nn
	%
	&&+\left(\frac{2 (D-2) G(2,3,4;1,3,4)^2}{(D-4) G(1,2,3,4;1,2,3,4)^2}+\frac{2 G(1,3,4) G(2,3,4)}{G(1,2,3,4;1,2,3,4)^2}\right) f_2 f_1\nn
	&&+\left(-\frac{2 (D-2) G(2,3,4;1,2,4) G(2,3,4;1,3,4)}{(D-4) G(1,2,3,4;1,2,3,4)^2}-\frac{2 G(1,3,4;1,2,4) G(2,3,4)}{G(1,2,3,4;1,2,3,4)^2}\right) f_3 f_1\nn
	&&+\left(\frac{2 (D-2) G(2,3,4;1,2,3) G(2,3,4;1,3,4)}{(D-4) G(1,2,3,4;1,2,3,4)^2}+\frac{2 G(1,3,4;1,2,3) G(2,3,4)}{G(1,2,3,4;1,2,3,4)^2}\right) f_4 f_1\nn
	&&+\left(-\frac{2 G(1,3,4;1,2,4) G(2,3,4;1,2,4)}{G(1,2,3,4;1,2,3,4)^2}-\frac{2 G(1,2,4;1,2,4) G(2,3,4;1,3,4)}{(D-4) G(1,2,3,4;1,2,3,4)^2}\right) f_3^2\nn
	&&+\left(-\frac{2 G(1,3,4;1,2,3) G(2,3,4;1,2,3)}{G(1,2,3,4;1,2,3,4)^2}-\frac{2 G(1,2,3) G(2,3,4;1,3,4)}{(D-4) G(1,2,3,4;1,2,3,4)^2}\right) f_4^2\nn
	&&+\left(\frac{2 G(1,3,4) G(2,3,4;1,2,4)}{G(1,2,3,4;1,2,3,4)^2}+\frac{2 (D-2) G(1,3,4;1,2,4) G(2,3,4;1,3,4)}{(D-4) G(1,2,3,4;1,2,3,4)^2}\right) f_2 f_3\nn
	&&+\left(-\frac{2 G(1,3,4) G(2,3,4;1,2,3)}{G(1,2,3,4;1,2,3,4)^2}-\frac{2 (D-2) G(1,3,4;1,2,3) G(2,3,4;1,3,4)}{(D-4) G(1,2,3,4;1,2,3,4)^2}\right) f_2 f_4\nn
	&&+\left(\frac{2 G(1,3,4;1,2,4) G(2,3,4;1,2,3)}{G(1,2,3,4;1,2,3,4)^2}+\frac{2 G(1,3,4;1,2,3) G(2,3,4;1,2,4)}{G(1,2,3,4;1,2,3,4)^2}\right)f_3 f_4\nn
	&&+\frac{4 G(1,2,4;1,2,3) G(2,3,4;1,3,4)}{(D-4) G(1,2,3,4;1,2,3,4)^2} f_3 f_4-\frac{2 (D-3) G(1,3,4) G(2,3,4;1,3,4) f_2^2}{(D-4) G(1,2,3,4;1,2,3,4)^2}.\nn
	\eea
	Other expansion coefficients can be got by using permutation symmetry:
	\bea
	c_{0,2,0,0}^{(0,1,2,3,4)}(2)&=&\left.c_{2,0,0,0}^{(0,1,2,3,4)}(2)\right\vert_{1\leftrightarrow 2},\nn
	c_{0,0,2,0}^{(0,1,2,3,4)}(2)&=&\left.c_{2,0,0,0}^{(0,1,2,3,4)}(2)\right\vert_{1\leftrightarrow 3},\nn
	c_{0,0,0,2}^{(0,1,2,3,4)}(2)&=&\left.c_{2,0,0,0}^{(0,1,2,3,4)}(2)\right\vert_{1\leftrightarrow 4},\nn
	c_{1,0,1,0}^{(0,1,2,3,4)}(2)&=&\left.c_{1,1,0,0}^{(0,1,2,3,4)}(2)\right\vert_{2\leftrightarrow 3},\nn
	c_{0,1,1,0}^{(0,1,2,3,4)}(2)&=&\left.c_{1,1,0,0}^{(0,1,2,3,4)}(2)\right\vert_{1\leftrightarrow 2},\nn
	c_{0,1,0,1}^{(0,1,2,3,4)}(2)&=&\left.c_{1,0,0,1}^{(0,1,2,3,4)}(2)\right\vert_{1\leftrightarrow 2},\nn
	c_{0,0,1,1}^{(0,1,2,3,4)}(2)&=&\left.c_{1,0,0,1}^{(0,1,2,3,4)}(2)\right\vert_{1\leftrightarrow 3},\nn
	c_{1,0,0,1}^{(0,1,2,3,4)}(2)&=&\left.c_{1,0,0,1}^{(0,1,2,3,4)}(2)\right\vert_{1\leftrightarrow 4}.
	\eea
\end{itemize}

\bibliographystyle{JHEP}
\bibliography{reference}

\providecommand{\href}[2]{#2}\begingroup\raggedright\begin{thebibliography}{10}

\bibitem{Feng:2021enk}
B.~Feng, T.~Li, and X.~Li, {\it {Analytic tadpole coefficients of one-loop
  integrals}},  {\em JHEP} {\bf 09} (2021) 081,
  [\href{http://arxiv.org/abs/2107.03744}{{\tt arXiv:2107.03744}}].

\bibitem{Brown:1952eu}
L.~M. Brown and R.~P. Feynman, {\it {Radiative corrections to Compton
  scattering}},  {\em Phys. Rev.} {\bf 85} (1952) 231--244.

\bibitem{Melrose:1965kb}
D.~B. Melrose, {\it {Reduction of Feynman diagrams}},  {\em Nuovo Cim.} {\bf
  40} (1965) 181--213.

\bibitem{Passarino:1978jh}
G.~Passarino and M.~J.~G. Veltman, {\it {One Loop Corrections for e+ e-
  Annihilation Into mu+ mu- in the Weinberg Model}},  {\em Nucl. Phys. B} {\bf
  160} (1979) 151--207.

\bibitem{tHooft:1978jhc}
G.~'t~Hooft and M.~J.~G. Veltman, {\it {Scalar One Loop Integrals}},  {\em
  Nucl. Phys. B} {\bf 153} (1979) 365--401.

\bibitem{vanNeerven:1983vr}
W.~L. van Neerven and J.~A.~M. Vermaseren, {\it {LARGE LOOP INTEGRALS}},  {\em
  Phys. Lett. B} {\bf 137} (1984) 241--244.

\bibitem{Stuart:1987tt}
R.~G. Stuart, {\it {Algebraic Reduction of One Loop Feynman Diagrams to Scalar
  Integrals}},  {\em Comput. Phys. Commun.} {\bf 48} (1988) 367--389.

\bibitem{vanOldenborgh:1989wn}
G.~J. van Oldenborgh and J.~A.~M. Vermaseren, {\it {New Algorithms for One Loop
  Integrals}},  {\em Z. Phys. C} {\bf 46} (1990) 425--438.

\bibitem{Bern:1992em}
Z.~Bern, L.~J. Dixon, and D.~A. Kosower, {\it {Dimensionally regulated one loop
  integrals}},  {\em Phys. Lett. B} {\bf 302} (1993) 299--308,
  [\href{http://arxiv.org/abs/hep-ph/9212308}{{\tt hep-ph/9212308}}]. [Erratum:
  Phys.Lett.B 318, 649 (1993)].

\bibitem{Bern:1993kr}
Z.~Bern, L.~J. Dixon, and D.~A. Kosower, {\it {Dimensionally regulated pentagon
  integrals}},  {\em Nucl. Phys. B} {\bf 412} (1994) 751--816,
  [\href{http://arxiv.org/abs/hep-ph/9306240}{{\tt hep-ph/9306240}}].

\bibitem{Fleischer:1999hq}
J.~Fleischer, F.~Jegerlehner, and O.~V. Tarasov, {\it {Algebraic reduction of
  one loop Feynman graph amplitudes}},  {\em Nucl. Phys. B} {\bf 566} (2000)
  423--440, [\href{http://arxiv.org/abs/hep-ph/9907327}{{\tt hep-ph/9907327}}].

\bibitem{Binoth:1999sp}
T.~Binoth, J.~P. Guillet, and G.~Heinrich, {\it {Reduction formalism for
  dimensionally regulated one loop N point integrals}},  {\em Nucl. Phys. B}
  {\bf 572} (2000) 361--386, [\href{http://arxiv.org/abs/hep-ph/9911342}{{\tt
  hep-ph/9911342}}].

\bibitem{Denner:2002ii}
A.~Denner and S.~Dittmaier, {\it {Reduction of one loop tensor five point
  integrals}},  {\em Nucl. Phys. B} {\bf 658} (2003) 175--202,
  [\href{http://arxiv.org/abs/hep-ph/0212259}{{\tt hep-ph/0212259}}].

\bibitem{Duplancic:2003tv}
G.~Duplancic and B.~Nizic, {\it {Reduction method for dimensionally regulated
  one loop N point Feynman integrals}},  {\em Eur. Phys. J. C} {\bf 35} (2004)
  105--118, [\href{http://arxiv.org/abs/hep-ph/0303184}{{\tt hep-ph/0303184}}].

\bibitem{Denner:2005nn}
A.~Denner and S.~Dittmaier, {\it {Reduction schemes for one-loop tensor
  integrals}},  {\em Nucl. Phys. B} {\bf 734} (2006) 62--115,
  [\href{http://arxiv.org/abs/hep-ph/0509141}{{\tt hep-ph/0509141}}].

\bibitem{Ellis:2007qk}
R.~K. Ellis and G.~Zanderighi, {\it {Scalar one-loop integrals for QCD}},  {\em
  JHEP} {\bf 02} (2008) 002, [\href{http://arxiv.org/abs/0712.1851}{{\tt
  arXiv:0712.1851}}].

\bibitem{Ossola_2007}
G.~Ossola, C.~G. Papadopoulos, and R.~Pittau, {\it Reducing full one-loop
  amplitudes to scalar integrals at the integrand level},  {\em Nuclear Physics
  B} {\bf 763} (Feb, 2007) 147--169.

\bibitem{Bern:1994cg}
Z.~Bern, L.~J. Dixon, D.~C. Dunbar, and D.~A. Kosower, {\it {Fusing gauge
  theory tree amplitudes into loop amplitudes}},  {\em Nucl. Phys. B} {\bf 435}
  (1995) 59--101, [\href{http://arxiv.org/abs/hep-ph/9409265}{{\tt
  hep-ph/9409265}}].

\bibitem{Chetyrkin:1981qh}
K.~G. Chetyrkin and F.~V. Tkachov, {\it {Integration by Parts: The Algorithm to
  Calculate beta Functions in 4 Loops}},  {\em Nucl. Phys. B} {\bf 192} (1981)
  159--204.

\bibitem{Tkachov:1981wb}
F.~V. Tkachov, {\it {A Theorem on Analytical Calculability of Four Loop
  Renormalization Group Functions}},  {\em Phys. Lett. B} {\bf 100} (1981)
  65--68.

\bibitem{Ossola:2006us}
G.~Ossola, C.~G. Papadopoulos, and R.~Pittau, {\it {Reducing full one-loop
  amplitudes to scalar integrals at the integrand level}},  {\em Nucl. Phys. B}
  {\bf 763} (2007) 147--169, [\href{http://arxiv.org/abs/hep-ph/0609007}{{\tt
  hep-ph/0609007}}].

\bibitem{Ossola:2007bb}
G.~Ossola, C.~G. Papadopoulos, and R.~Pittau, {\it {Numerical evaluation of
  six-photon amplitudes}},  {\em JHEP} {\bf 07} (2007) 085,
  [\href{http://arxiv.org/abs/0704.1271}{{\tt arXiv:0704.1271}}].

\bibitem{Ellis:2007br}
R.~K. Ellis, W.~T. Giele, and Z.~Kunszt, {\it {A Numerical Unitarity Formalism
  for Evaluating One-Loop Amplitudes}},  {\em JHEP} {\bf 03} (2008) 003,
  [\href{http://arxiv.org/abs/0708.2398}{{\tt arXiv:0708.2398}}].

\bibitem{Bern:1994zx}
Z.~Bern, L.~J. Dixon, D.~C. Dunbar, and D.~A. Kosower, {\it {One loop n point
  gauge theory amplitudes, unitarity and collinear limits}},  {\em Nucl. Phys.
  B} {\bf 425} (1994) 217--260,
  [\href{http://arxiv.org/abs/hep-ph/9403226}{{\tt hep-ph/9403226}}].

\bibitem{Britto:2004nc}
R.~Britto, F.~Cachazo, and B.~Feng, {\it {Generalized unitarity and one-loop
  amplitudes in N=4 super-Yang-Mills}},  {\em Nucl. Phys. B} {\bf 725} (2005)
  275--305, [\href{http://arxiv.org/abs/hep-th/0412103}{{\tt hep-th/0412103}}].

\bibitem{Britto:2005ha}
R.~Britto, E.~Buchbinder, F.~Cachazo, and B.~Feng, {\it {One-loop amplitudes of
  gluons in SQCD}},  {\em Phys. Rev. D} {\bf 72} (2005) 065012,
  [\href{http://arxiv.org/abs/hep-ph/0503132}{{\tt hep-ph/0503132}}].

\bibitem{Campbell:1996zw}
J.~M. Campbell, E.~W.~N. Glover, and D.~J. Miller, {\it {One loop tensor
  integrals in dimensional regularization}},  {\em Nucl. Phys. B} {\bf 498}
  (1997) 397--442, [\href{http://arxiv.org/abs/hep-ph/9612413}{{\tt
  hep-ph/9612413}}].

\bibitem{Bern:1997sc}
Z.~Bern, L.~J. Dixon, and D.~A. Kosower, {\it {One loop amplitudes for e+ e- to
  four partons}},  {\em Nucl. Phys. B} {\bf 513} (1998) 3--86,
  [\href{http://arxiv.org/abs/hep-ph/9708239}{{\tt hep-ph/9708239}}].

\bibitem{Anastasiou:2006gt}
C.~Anastasiou, R.~Britto, B.~Feng, Z.~Kunszt, and P.~Mastrolia, {\it {Unitarity
  cuts and Reduction to master integrals in d dimensions for one-loop
  amplitudes}},  {\em JHEP} {\bf 03} (2007) 111,
  [\href{http://arxiv.org/abs/hep-ph/0612277}{{\tt hep-ph/0612277}}].

\bibitem{Britto:2010um}
R.~Britto and E.~Mirabella, {\it {Single Cut Integration}},  {\em JHEP} {\bf
  01} (2011) 135, [\href{http://arxiv.org/abs/1011.2344}{{\tt
  arXiv:1011.2344}}].

\bibitem{Anastasiou:2006jv}
C.~Anastasiou, R.~Britto, B.~Feng, Z.~Kunszt, and P.~Mastrolia, {\it
  {D-dimensional unitarity cut method}},  {\em Phys. Lett. B} {\bf 645} (2007)
  213--216, [\href{http://arxiv.org/abs/hep-ph/0609191}{{\tt hep-ph/0609191}}].

\bibitem{Britto:2006fc}
R.~Britto and B.~Feng, {\it {Unitarity cuts with massive propagators and
  algebraic expressions for coefficients}},  {\em Phys. Rev. D} {\bf 75} (2007)
  105006, [\href{http://arxiv.org/abs/hep-ph/0612089}{{\tt hep-ph/0612089}}].

\bibitem{Britto:2007tt}
R.~Britto and B.~Feng, {\it {Integral coefficients for one-loop amplitudes}},
  {\em JHEP} {\bf 02} (2008) 095, [\href{http://arxiv.org/abs/0711.4284}{{\tt
  arXiv:0711.4284}}].

\bibitem{Britto:2008vq}
R.~Britto, B.~Feng, and P.~Mastrolia, {\it {Closed-Form Decomposition of
  One-Loop Massive Amplitudes}},  {\em Phys. Rev. D} {\bf 78} (2008) 025031,
  [\href{http://arxiv.org/abs/0803.1989}{{\tt arXiv:0803.1989}}].

\bibitem{Britto:2008sw}
R.~Britto, B.~Feng, and G.~Yang, {\it {Polynomial Structures in One-Loop
  Amplitudes}},  {\em JHEP} {\bf 09} (2008) 089,
  [\href{http://arxiv.org/abs/0803.3147}{{\tt arXiv:0803.3147}}].

\bibitem{Feng:2013sa}
B.~Feng and H.~Wang, {\it {Analytic structure of one-loop coefficients}},  {\em
  JHEP} {\bf 05} (2013) 104, [\href{http://arxiv.org/abs/1301.7510}{{\tt
  arXiv:1301.7510}}].

\bibitem{Britto:2004ap}
R.~Britto, F.~Cachazo, and B.~Feng, {\it {New recursion relations for tree
  amplitudes of gluons}},  {\em Nucl. Phys. B} {\bf 715} (2005) 499--522,
  [\href{http://arxiv.org/abs/hep-th/0412308}{{\tt hep-th/0412308}}].

\bibitem{Britto:2005fq}
R.~Britto, F.~Cachazo, B.~Feng, and E.~Witten, {\it {Direct proof of tree-level
  recursion relation in Yang-Mills theory}},  {\em Phys. Rev. Lett.} {\bf 94}
  (2005) 181602, [\href{http://arxiv.org/abs/hep-th/0501052}{{\tt
  hep-th/0501052}}].

\bibitem{Smirnov:2008iw}
A.~V. Smirnov, {\it {Algorithm FIRE -- Feynman Integral REduction}},  {\em
  JHEP} {\bf 10} (2008) 107, [\href{http://arxiv.org/abs/0807.3243}{{\tt
  arXiv:0807.3243}}].

\bibitem{Smirnov:2014hma}
A.~V. Smirnov, {\it {FIRE5: a C++ implementation of Feynman Integral
  REduction}},  {\em Comput. Phys. Commun.} {\bf 189} (2015) 182--191,
  [\href{http://arxiv.org/abs/1408.2372}{{\tt arXiv:1408.2372}}].

\bibitem{Smirnov:2013dia}
A.~V. Smirnov and V.~A. Smirnov, {\it {FIRE4, LiteRed and accompanying tools to
  solve integration by parts relations}},  {\em Comput. Phys. Commun.} {\bf
  184} (2013) 2820--2827, [\href{http://arxiv.org/abs/1302.5885}{{\tt
  arXiv:1302.5885}}].

\bibitem{Smirnov:2019qkx}
A.~V. Smirnov and F.~S. Chuharev, {\it {FIRE6: Feynman Integral REduction with
  Modular Arithmetic}},  {\em Comput. Phys. Commun.} {\bf 247} (2020) 106877,
  [\href{http://arxiv.org/abs/1901.07808}{{\tt arXiv:1901.07808}}].

\bibitem{Lee:2013mka}
R.~N. Lee, {\it {LiteRed 1.4: a powerful tool for reduction of multiloop
  integrals}},  {\em J. Phys. Conf. Ser.} {\bf 523} (2014) 012059,
  [\href{http://arxiv.org/abs/1310.1145}{{\tt arXiv:1310.1145}}].

\end{thebibliography}\endgroup

\end{document}